\documentclass[a4paper,fleqn,usenatbib,useAMS]{mnras}

\usepackage{graphicx}	
\usepackage{amsmath}	
\usepackage{mathtools}
\usepackage{color}

\newcommand{\fracb}[2]{\left(\frac{#1}{#2}\right)}
\newcommand{\sign}{\text{sign}}

\usepackage[T1]{fontenc}
\usepackage{ae,aecompl}

\usepackage{newtxtext,newtxmath}

\title[2D RMHD Simulations of the K-S Instability]
{2D Relativistic MHD Simulations of the Kruskal-Schwarzschild Instability in a Relativistic Striped Wind}

\author[Gill, Granot, \& Lyubarsky 2017]{Ramandeep Gill,$^{1,2}$\thanks{Contact e-mail:
\href{mailto:rsgill.rg@gmail.com}{rsgill.rg@gmail.com}}
Jonathan Granot,$^{1}$\thanks{Contact e-mail:
\href{mailto:granot@openu.ac.il}{granot@openu.ac.il}}
and Yuri Lyubarsky$^{2}$\thanks{Contact e-mail:
\href{mailto:lyub@bgu.ac.il}{lyub@bgu.ac.il}}
\\
$^{1}$Department of Natural Sciences, The Open University of Israel, 
1 University Road, PO Box 808, Raanana 4353701, Israel \\
$^{2}$Physics Department, Ben-Gurion University, P.O.B. 653, 
Beer-Sheva 84105, Israel}

\date{Last updated; in original form}

\pubyear{2017}

\begin{document}
\label{firstpage}
\pagerange{\pageref{firstpage}--\pageref{lastpage}}
\maketitle

\begin{abstract}
We study the linear and non-linear development of the Kruskal-Schwarzchild Instability in a 
relativisitically expanding striped wind. This instability is the generalization of Rayleigh-Taylor instability in the presence of a magnetic field. It has been suggested to produce a self-sustained 
acceleration mechanism in strongly magnetized outflows found in active galactic nuclei, gamma-ray 
bursts, and micro-quasars. The instability leads to magnetic reconnection, but in contrast with steady-state Sweet-Parker reconnection, the dissipation rate is not limited by the current layer's small aspect ratio. We performed two-dimensional (2D) relativistic magneto-hydrodynamic (RMHD) simulations featuring two cold and highly magnetized ($1\leq\sigma\leq10^{3}$) plasma layers with an anti-parallel magnetic field separated by a thin layer of relativistically hot plasma with a local effective gravity induced by the outflow's acceleration. Our simulations 
show how the heavier relativistically hot plasma in the reconnecting layer drips out and allows oppositely 
oriented magnetic field lines to reconnect. The instability's growth rate in the linear regime matches the predictions of linear stability analysis. We find turbulence rather than an ordered bulk flow near the reconnection region, with turbulent velocities
up to $\sim0.1$c, largely independent of model 
parameters. However, the magnetic energy dissipation rate is found to be much slower, corresponding to an effective ordered bulk velocity inflow into the reconnection region $v_{\rm in}=\beta_{\rm in}c$, of $10^{-3}\lesssim\beta_{\rm in}\lesssim 5\times10^{-3}$. This occurs due 
to the slow evacuation of hot plasma from the current layer, largely because of the Kelvin-Helmholtz instability experienced by the dripping plasma. 3D RMHD simulations are needed to further investigate the non-linear regime.
\end{abstract}


\section{Introduction}

Relativistic outflows are ubiquitous in the Universe. They are usually collimated into narrow jets,  which are either 
observed directly, namely in active galactic nuclei (AGN; e.g. jets in M87 and Cygnus A), or inferred indirectly 
from multi-wavelength observations in X-ray binaries i.e. micro-quasars \citep[e.g.][]{FBG04},
gamma-ray bursts \citep[GRBs; e.g.][]{KZ15}, and tidal disruption events \citep[e.g.][]{KP12}. Relativistic outflows typically arise from accretion onto a rapidly spinning central compact object, such as a black hole (BH) or a neutron star (NS), which leads to the expulsion of matter and entrained magnetic fields at relativistic speeds. In pulsar winds and possibly in some GRBs the relativistic outflow is powered by the rotational energy of a rapidly rotating neutron star central source, rather than accretion. The launching, collimation, and acceleration of  relativistic outflows to high bulk Lorentz factors $\Gamma \gg 1$ in such a variety of systems is an active area of research.

The composition of relativistic jets or outflows in the different astrophysical sources, and in particular their degree of magnetization, is highly uncertain and of great interest. Pulsar winds are almost certainly Poynting flux dominated near the central source, and most likely so are the outflows from active galactic nuclei (AGN) and tidal disruption events (TDEs) of a star by a super-massive black hole. In AGN and TDEs, since the central accreting black hole is super-massive, then even close to it the Thomson optical depth $\tau_T$ may not be high enough for thermal acceleration by radiation pressure -- the main competition to magnetic acceleration -- to work efficiently \citep[e.g.][]{Ghisellini2011}. In GRBs or micro-quasars, however, thermal acceleration could also work ($\tau_T\gg1$ is possible, or even likely), and the dominant acceleration mechanism is less clear. Nonetheless, there is a growing consensus that such outflows are launched hydro-magnetically with the magnetic
fields playing a critical role \citep[see for e.g. reviews by][]{Spruit2010,PHG12,Granot+15}.

Here we consider outflows that are at least initially Poynting flux dominated. One of the most important open questions about such outflows that start out
highly magnetized near the central source is how they convert most of
their initial electromagnetic energy to other forms, namely bulk
kinetic energy and the energy in the random motions of the particles,
which also produce the radiation we observe from these
sources (i.e. the outflow acceleration, energy dissipation, particle acceleration and radiation). Observations of the relevant sources, such as AGN, GRBs or pulsar
wind nebulae, suggest that the outflow magnetization $\sigma$ (the Poynting-to-matter energy flux ratio) is rather low at
large distances from the source. This is known as the $\sigma$
problem, namely how to transform from $\sigma\gg 1$ near the source to
$\sigma\ll 1$ very far from the source.

Poynting flux dominated outflows are often treated under the simplifying assumptions of ideal MHD, axi-symmetry and steady-state. However, under these conditions it is very hard to achieve
$\sigma < 1$ (or $\sigma\ll 1$) far from the source that would enable
efficient energy dissipation in internal
shocks~\citep{Kom09,Lyub09,Lyub10a,Tchek+08,Tchek+09,Tchek10}, where the acceleration requires external pressure confinement and is tightly coupled to the collimation of the jet. While this process could lead to $\sigma\approx 1$ this requires rather restrictive conditions. 

Alternatively, the
non-axi-symmetric kink instability could randomize the direction of
the magnetic field, causing it to behave more like a fluid and
enhancing magnetic reconnection, which both increase the acceleration
and help lower the magnetization~\citep{Lyub92,Eichler93,Spruit+97,Begelman98,GS06,BT16}. 
However, such a global MHD instability could develop only if the proper Alfv{\'e}n (lateral) crossing time is less than the propagation time, which implies $\Gamma\theta_{\rm jet} < 1$, where $\Gamma$ is the jet Lorentz factor and $\theta_{\rm jet}$ is its half-opening angle. This condition is quite restrictive, e.g. it could hardly be fulfilled in GRBs \citep[e.g.][]{Tchek10}. Moreover, even if the kink instability develops, it is still not clear whether the flow is disrupted or simply helically distorted.

An efficient conversion of electromagnetic energy into kinetic and thermal energy of the plasma is possible in impulsive flows that have a strong time variability \citep{GKS11,Lyut11}. The maximal Lorentz factor $\Gamma$ and minimal magnetization $\sigma$ that can be reached by a single thick shell (of a few tens of light seconds, comparable to the duration of a long GRB) is somewhat limited due to the interaction with the external medium \citep{Levinson10,Granot12a}. This may be alleviated if the outflow consists of many thinner, well separated sub-shells \citep{Granot12a,Komissarov12}, where even if the collisions between these sub-shells as they expand radially start at $\sigma\gg1$ then gradual dissipation and subsequent acceleration can still occur via multiple passages of weak shocks.

An alternative option that we will focus on in this work involves magnetic energy dissipation in a striped wind. The prime example of a striped wind is a pulsar wind, where the pulsar acts as an oblique rotator with misaligned rotation and magnetic field symmetry axes, where the magnetic field of the outflowing MHD wind in the equatorial belt switches its polarity twice in each rotation period. 
The structure of the magnetic 
field advected at velocity $v$ with the particle outflow is that of a \textit{striped}-wind \citep{Michel71,Coroniti90,Michel94,LK01}, 
with field lines reversing polarity over a lab-frame length scale of $\lambda_B \approx \pi v/\Omega$. The polarity reversal 
between columns of magnetic field is marked by the presence of a current sheet towards which the magnetized fluid flows at a 
fraction of the Alfv\'{e}n speed in the fluid-frame. Magnetic reconnection in these current layers 
helps accelerate the flow and heats up particles creating a relativistically hot plasma in these current layers.

In pulsar winds a striped-wind arises naturally and magnetic field dissipation has been shown to be the main energy conversion mechanism there \citep{LK01,KS03,PL07,SS11}. Moreover, a broadly similar magnetic field configuration in the outflow may arise from accretion onto a black hole, due to stochastic flipping of the magnetic field polarity, possibly due to instabilities in the accretion disk \citep[][]{MU12}. This would result in a striped wind with shells of correspondingly random width and magnetic field polarity. Magnetic energy dissipation in a striped wind has received particular attention in the study of relativistic 
outflows ($\Gamma \gtrsim 100$) in GRBs, where the central engine 
(CE; a BH or a fast-rotating magnetar) launches a Poynting flux dominated outflow that suffers magnetic reconnection at a 
radial distance $r\sim10^{12}-10^{14}~{\rm cm}$ from the CE and produces the $\sim200~{\rm keV}-{\rm MeV}$ gamma-ray emission 
in the prompt-phase \citep[e.g.][]{DS02,Giannios08}. In general, magnetic reconnection that does not necessarily 
result from having a striped-wind configuration has also been invoked in many works over the traditional internal-shock scenario 
to explain the prompt GRB emission due to it being more efficient in strongly magnetized flows 
\citep[e.g.][]{Thompson94,LB03,GS06,Lyut06,ZY11}.

Astrophysical plasmas near compact objects are inherently relativistic and collisionless \citep[e.g.][]{LL13}, 
and they require some source of anomalous resistivity in order for reconnection to proceed. The rate of reconnection 
is set by the inflow velocity $v_{\rm in}=\beta_{\rm in}c$ of the magnetized fluid into the current layer. In fast reconnection models 
\citep[e.g.][]{Petschek64} it can be a fraction of the Alfv\'{e}n speed, which approaches the speed of light in 
Poynting flux dominated outflows. 
It was shown by \citet{Lyubarsky05} and then confirmed using numerical simulations 
by e.g. \citet{WY06,Guo+15} that $\beta_{\rm in}\sim0.1$ when magnetic reconnection occurs in the relativistic Petschek regime.

It is worth noting that 
in most analytic and numerical works on reconnection the plasma is allowed to freely stream out of the region surrounding the X-point along the reconnection layer.
In more realistic configuration featuring multiple X-points along the reconnection layer the hot plasma that is 
produced by the reconnection accumulates in the region between X-points and its dynamical effect can eventually slow down or even prevent further reconnection. This motivated search for mechanisms that could evacuate the hot plasma 
from the reconnection layer, which as a result would allow further reconnection and increase the reconnection rate.


This motivated the suggestion that
magnetic reconnection in Poynting flux dominated relativistic outflows with a striped wind structure can 
be facilitated by the Kruskal-Schwarzschild instability (KSI) of the current sheet \citep{L10}. This is an analog of the Rayleigh-Taylor instability (RTI) in strongly magnetized flows \citep[e.g][]{LG13}. It was shown that 
as the flow accelerates the current layer feels an effective (comoving) gravity $g = c^2d\Gamma(r)/dr$ in the opposite direction. 
Since the enthalpy density of the relativistically hot plasma in the current layer is larger than that of the cold magnetized 
plasma below it, the current sheet becomes susceptible to the KSI just like an interface between a lighter 
fluid below a heavier one would be to the RTI. As the hot plasma drips out of the reconnection layer it allows further magnetic reconnection that produces more hot plasma and accelerates the flow, thus creating a positive feedback loop, and makes this instability self-sustained.
Therefore, it can potentially account both for the acceleration of the flow as well as for the energy dissipation, which leads to particle acceleration and the radiation that we observe.
%
Moreover, this mechanism completely avoids the limitations of the classical Sweet-Parker resistive reconnection model and may 
yield fast reconnection rates. This is achieved by the downward (in the direction of the effective gravity force) 
dripping of hot plasma from the current layer where its removal is otherwise limited by the narrow width of the current 
layer in the Sweet-Parker model. By conservation of mass, faster removal of hot plasma from the current layer allows a 
faster inflow rate of cold magnetized gas, and thus a higher reconnection rate. 


The focus of the present work is to quantitatively understand the structure and temporal evolution of the KSI using 2D relativistic 
MHD (RMHD) simulations. The primary goals are (i) to confirm the growth rate of the instability in the linear 
stage as calculated by \citet{L10} using a linear stability analysis, and (ii) to obtain the energy dissipation rate in the 
non-linear regime, which can only be done using MHD simulations. To this end, in \S\ref{sec:LSA} we briefly discuss the results of the 
linear theory and in \S\ref{sec:2D-sim} present the numerical method used for the 2D simulations. We present the results from the simulations 
in \S\ref{sec:results}, where we discuss perturbation modes with wavelength much larger and comparable to the width of the current layer. 
In \S\ref{sec:Rec-rate}, we infer the rate of reconnection from the rate of magnetic field dissipation in the simulated volume and discuss the role of the Kelvin-Helmholtz Instability (KHI) and buoyancy in limiting this rate in \S\ref{sec:buoyancy}. 
We discuss some implications of this work in \S\ref{sec:discussion}. We adopt Lorentz-Heaviside units and set the speed of light $c=1$ throughout \S\ref{sec:2D-sim} - \S\ref{sec:buoyancy}.

\begin{figure}
    \includegraphics[width=0.48\textwidth]{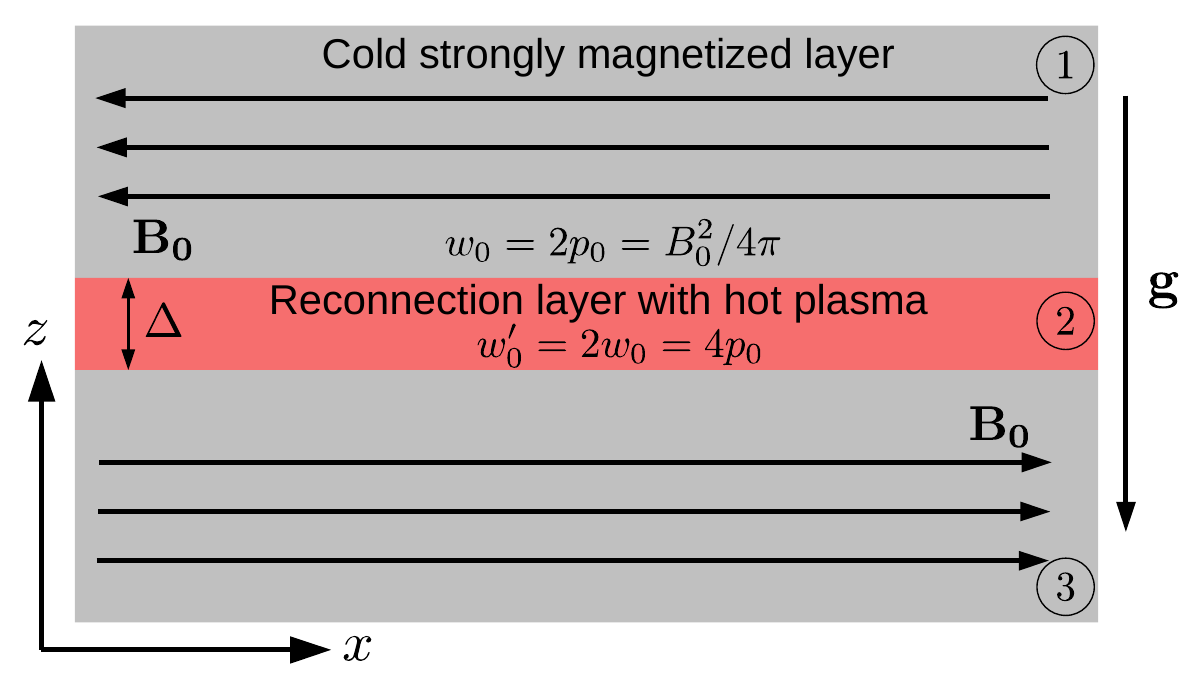}
    \caption{Striped wind with reconnecting layer in the comoving frame. 
    The hot reconnection layer (2) of width $\Delta$ and enthalpy density $w_0'$ 
    is surrounded by two (1 \& 3) cold but strongly magnetized layers with magnetic field $\mathbf{B_0}$. 
    Here the field lines are shown as completely anti-parallel in the two layers, however, more generally 
    they can also be slightly misaligned. The enthalpy density of the magnetized layers is $w_0 = w_0'/2$. 
    As the flow accelerates in the $\hat z$ direction, the layers feel an effective gravity 
    $\mathbf g=-g\hat z = -c^2d\Gamma(r)/dr\hat z$.}
    \label{fig:setup}
\end{figure}

\section{Linear Stability Analysis}\label{sec:LSA}
We consider a relativistically expanding outflow moving with bulk- Lorentz factor $\Gamma$. In the fluid-frame, we consider a non-magnetized, 
relativistically hot slab of plasma of width $\Delta$, surrounded by a cold but strongly magnetized plasma with $\textbf{B} = -\sign(z)B_0\hat x$. 
The magnetic field has opposite polarity on either side of the current layer (see Fig.~\ref{fig:setup}). The dynamical equations for the 
plasma, in the ideal MHD limit, follow directly from conservation laws \citep[e.g.][]{LL66}
\begin{equation}
\partial_\mu(\rho u^\mu) = 0,\quad\quad\partial_\mu T^{\mu\nu} = 0
\label{eq:conserv_laws}
\end{equation}
for $\mu = 0,1,2,3$, where $\partial_\mu = (\partial/\partial(ct),\nabla)$ is the four-derivative, 
$u^\mu=(\gamma c,\gamma\mathbf{v})$ is the fluid four-velocity. The velocity $\mathbf{v}$ and Lorentz factor $\gamma=[1-(v/c)^2]^{-1/2}$ are measured in the outflow's bulk frame, in which the simulation is performed. 
The stress-energy tensor receives contributions from 
both the plasma and electromagnetic components, $T^{\mu\nu} = T^{\mu\nu}_{\rm pl} + T^{\mu\nu}_{\rm em}$, such that \citep[e.g.][]{GKP10}
\begin{equation}
T^{\mu\nu} = \frac{wu^\mu u^\nu}{c^2}+p\eta^{\mu\nu} - b^\mu b^\nu~,
\end{equation}
where $\sqrt{4\pi}b^\mu = \left[\gamma(\mathbf{B}\cdot\mathbf{v})/c,\,\mathbf{B}/\gamma + \gamma(\mathbf{B}\cdot\mathbf{v})\mathbf{v}/c^2\right]$ is the magnetic field four-vector, and $\eta^{\mu\nu}=\rm{diag}(-1,1,1,1)$ is the 
Minkowski metric. The enthalpy density in the fluid rest frame in each layer is
\begin{equation}
w = \rho c^2 + \frac{\hat \gamma}{\hat \gamma - 1}p
\end{equation}
where $\rho$ and $p$ are the fluid-frame plasma mass density and pressure, and $\hat \gamma$ is the adiabatic index. In the strongly 
magnetized layer, the pressure is dominated by that of the magnetic field, for which $\hat\gamma\rightarrow2$ and $p_B = B^2/8\pi$. 

From Eq.~(\ref{eq:conserv_laws}), and keeping only the terms non-vanishing to first order in the perturbative expansion that follows, 
we find
\begin{eqnarray}
&&\frac{\partial\rho}{\partial t} + \nabla\cdot(\rho\mathbf{v}) = 0~, \\
&& w\frac{\partial\mathbf{v}}{\partial t} = -c^2\nabla p + 
\frac{(\mathbf{B}\cdot\nabla)\mathbf{B}}{4\pi} + w\mathbf{g}~.
\label{eq:momentum_conserv}
\end{eqnarray}
The $(\mathbf{B}\cdot\nabla)\mathbf{B}  /4\pi$ term represents the force due to magnetic tension, and 
$w\mathbf{g}$ is the effective gravity force felt by the fluid in the bulk-flow frame. As the flow accelerates in the 
$\hat z$-direction, the inertial acceleration is aligned in the opposite direction, $\mathbf{g} = -g\hat z$. These equations are 
further supplemented by the flux-freezing condition
\begin{equation}
\frac{\partial\mathbf{B}}{\partial t} = \nabla\times(\mathbf{v}\times\mathbf{B})~,
\end{equation}
and the equation of state
\begin{equation}
\frac{d}{dt}\fracb{p}{\rho^{\hat \gamma}} = 0~,
\end{equation}
which expresses the adiabatic condition \textcolor{black}{and only holds in the linear stage when no magnetic flux is destroyed by 
reconnection}. The \textit{zero}-th order equations, expressing the equilibrium state of the fluid, yield, 
$\partial \rho_0/\partial t=\partial p_0/\partial t=\partial w_0/\partial t = \partial \mathbf{B}_0/\partial t = 0$, and 
$\partial_xp_0 = \partial_yp_0 = 0$, with vertical pressure stratification condition due to the effective gravity
\begin{equation}
\partial_z p_0 = -\frac{w_0g}{c^2}~,
\end{equation}
which for a homogeneous density has the solution
\begin{equation}
p_0(z) = p_0(0)-w_0\frac{gz}{c^2} = p_0(0)\left[1+\mathcal{O}\left(\frac{gz}{c^2}\right)\right]~,
\end{equation}
suggesting that on length scales $z\ll c^2/g$, the pressure is also homogeneous. Furthermore, noting that in both the 
hot unmagnetized central layer and in the two cold but strongly magnetized layers $p_0\sim w_0$, the pressure stratification condition leads to the result 
\begin{equation}
\Delta \ll \frac{c^2}{g}\equiv L_{\rm dyn}~,
\label{eq:delta_condition}
\end{equation}
in order to have a small fractional change in the initial pressure across the central layer. Here $L_{\rm dyn}$ is the dynamical length below which the effective gravity $g$ is approximately constant and above which it may change significantly. This may be understood since $g=c^2d\Gamma(r)/dr \sim c^2\Gamma/r$ so that $c^2/g\sim r/\Gamma$ is of the order of the causal length in the radial direction. Defining the corresponding comoving dynamical time, $t_{\rm dyn}\equiv L_{\rm dyn}/c$, this corresponds to $gt_{\rm dyn}\sim c$, i.e. a Newtonian free-fall velocity of order $c$ is achieved over the dynamical time.

By considering small amplitude perturbations in the equilibrium quantities and adopting the ansatz that the 
perturbed quantities vary harmonically (e.g. $\rho_1 \propto \exp(i[ky-\omega t])$ where 
$k$ is the wavenumber and $\omega$ is the wave frequency), \citet{L10} \textcolor{black}{(see for full derivation of the linear growth rate)} 
obtained the following dispersion relation for modes orthogonal to the equilibrium magnetic field,
\begin{equation}
    \omega^2_\pm = \pm gk\left(5+\frac{4}{\tanh(k\Delta)}\right)^{-1/2}~.
    \label{eq:general-disp-rel}
\end{equation}
The growth rate of the instability is then given by $\eta = {\rm Im}(\omega_-)$, which asymptotes to the 
following in the small ($k\Delta\gg1$) and large ($k\Delta\ll1$) wavelength limits
\begin{equation}
\eta = \begin{cases}
    \sqrt{\dfrac{gk}{3}}~, & k\Delta\gg 1 \\
    \left(\dfrac{g}{2}\right)^{1/2}k^{3/4}\Delta^{1/4}~, & k\Delta\ll1~.
\end{cases}
\end{equation}

\begin{figure}
    \includegraphics[width=0.48\textwidth]{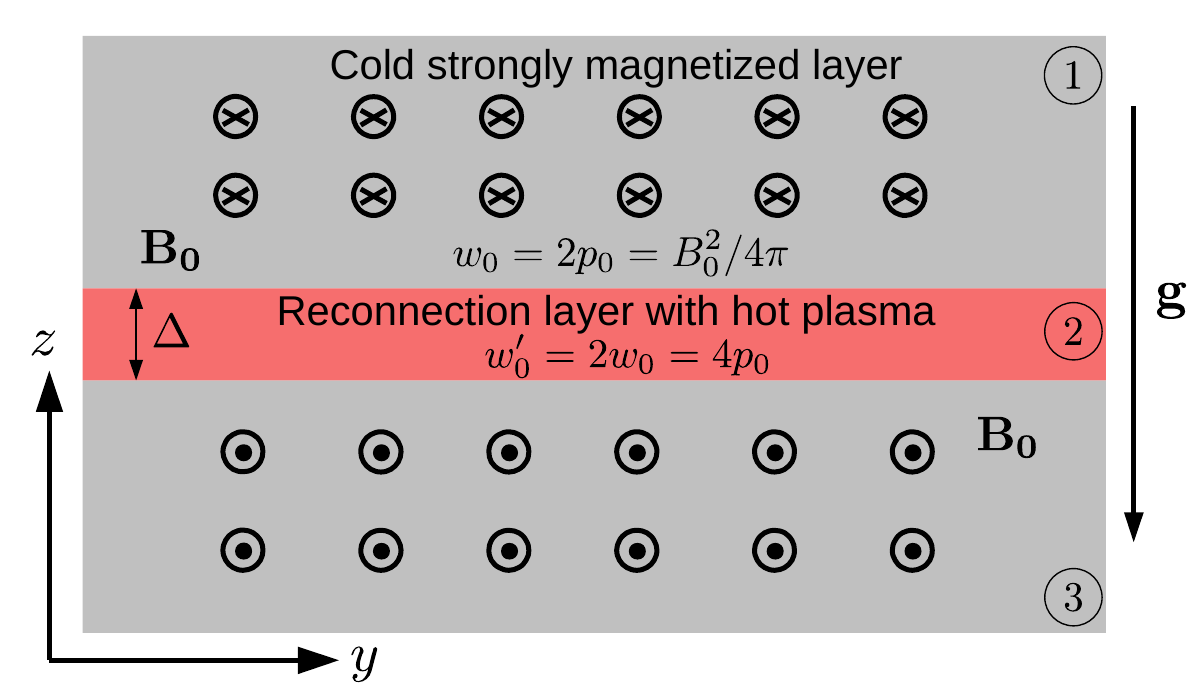}
    \caption{Setup of the simulation box, which is rotated here by $90^\circ$ with respect to the illustration of the 
    striped wind with a reconnecting layer as shown in Fig.~\ref{fig:setup}. The magnetic field lines go into the page 
    in region 1 and come out of the page in region 3.}
    \label{fig:sim-setup}
\end{figure}

\section{2D Numerical Simulations}\label{sec:2D-sim}

\begin{figure*}
    \centering
    \includegraphics[width=\textwidth]{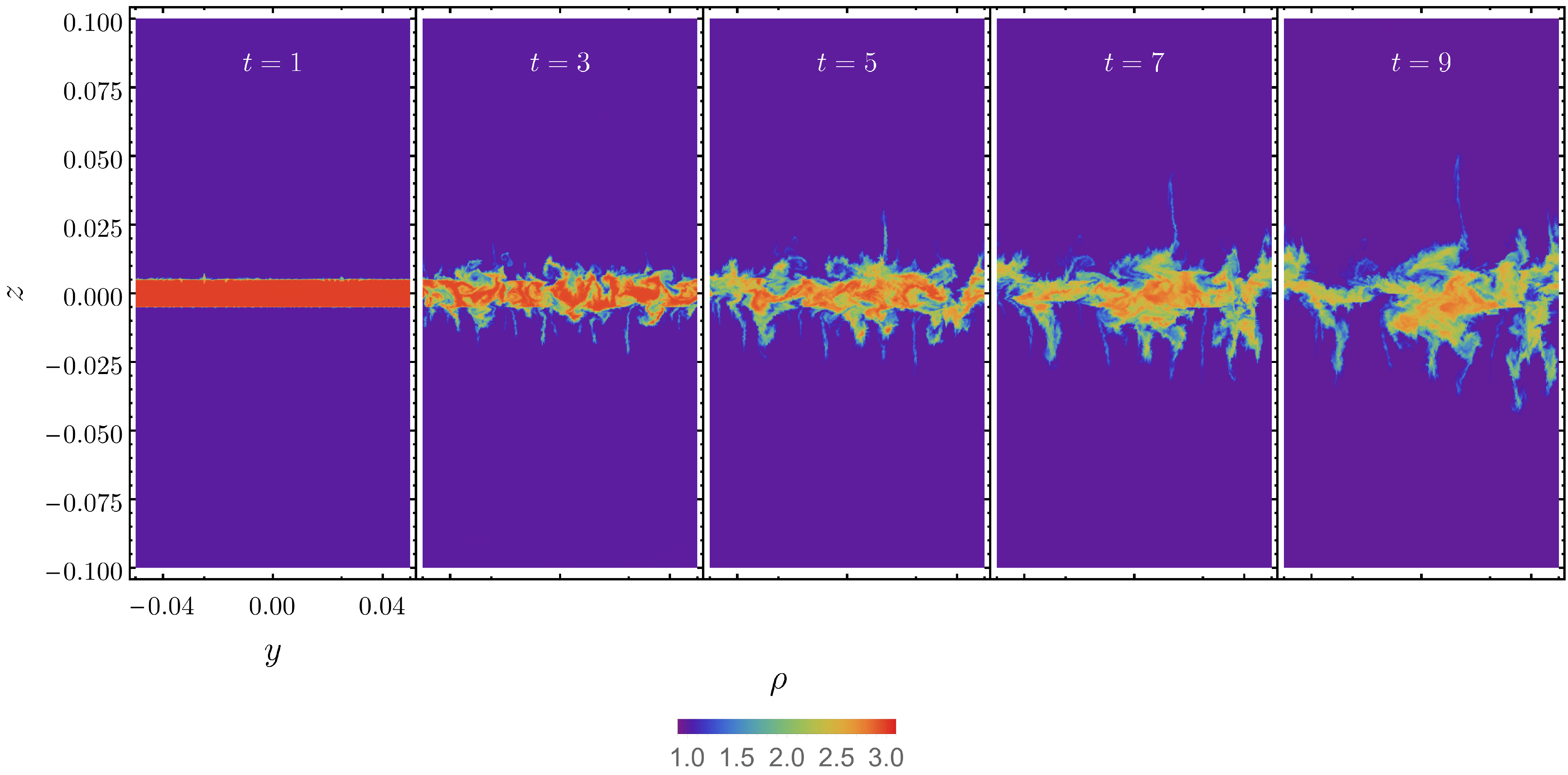}
    \caption{Development of the KSI from the linear to the non-linear stage, shown here using the 
    fluid mass density, in the single wavelength $(m_0=1)$ $\lambda_0 = 0.1$ case. The initial velocity perturbation 
    amplitude $\kappa_v = v_1/v_A = 10^{-6}$ of the Alfv{\'e}n speed in the magnetized layer. The effective 
    gravity points downwards and has magnitude $g = 0.01$, which gives a causal size of $z = g^{-1} = 100$ 
    over which pressure homogeneity is maintained. The magnetization $\sigma_0 = 10$ and 
    the density contrast between the cold magnetized and the hot unmagnetized layer is 
    $\rho_{0,h}/\rho_{0,c}\equiv\psi=3$ with $\rho_{0,c} = 1$.}
    \label{fig:2D-m1-g001}
\end{figure*}
\begin{figure*}
    \centering
    \includegraphics[width=0.45\textwidth]{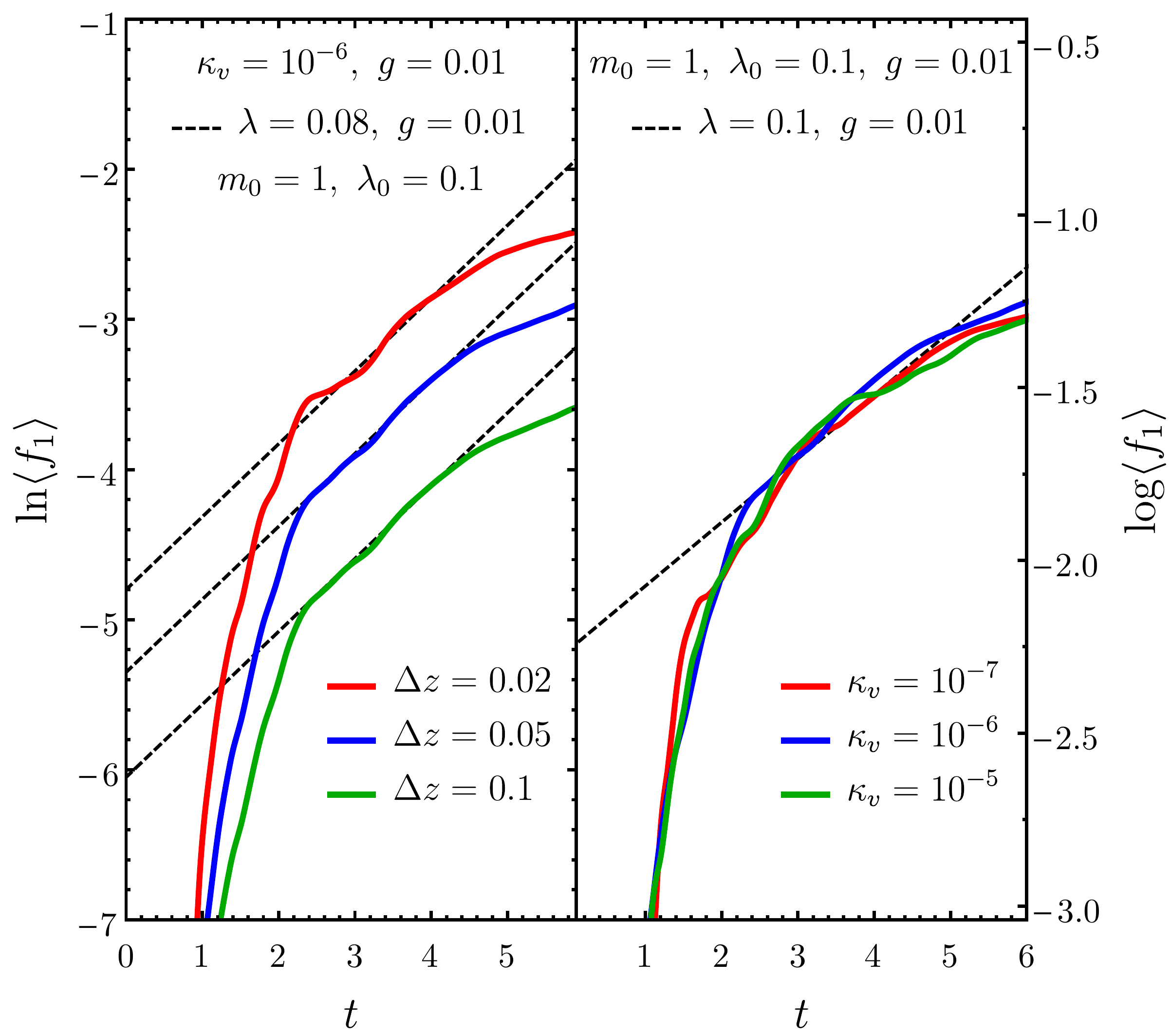}\hspace{1cm}
    \includegraphics[width=0.45\textwidth]{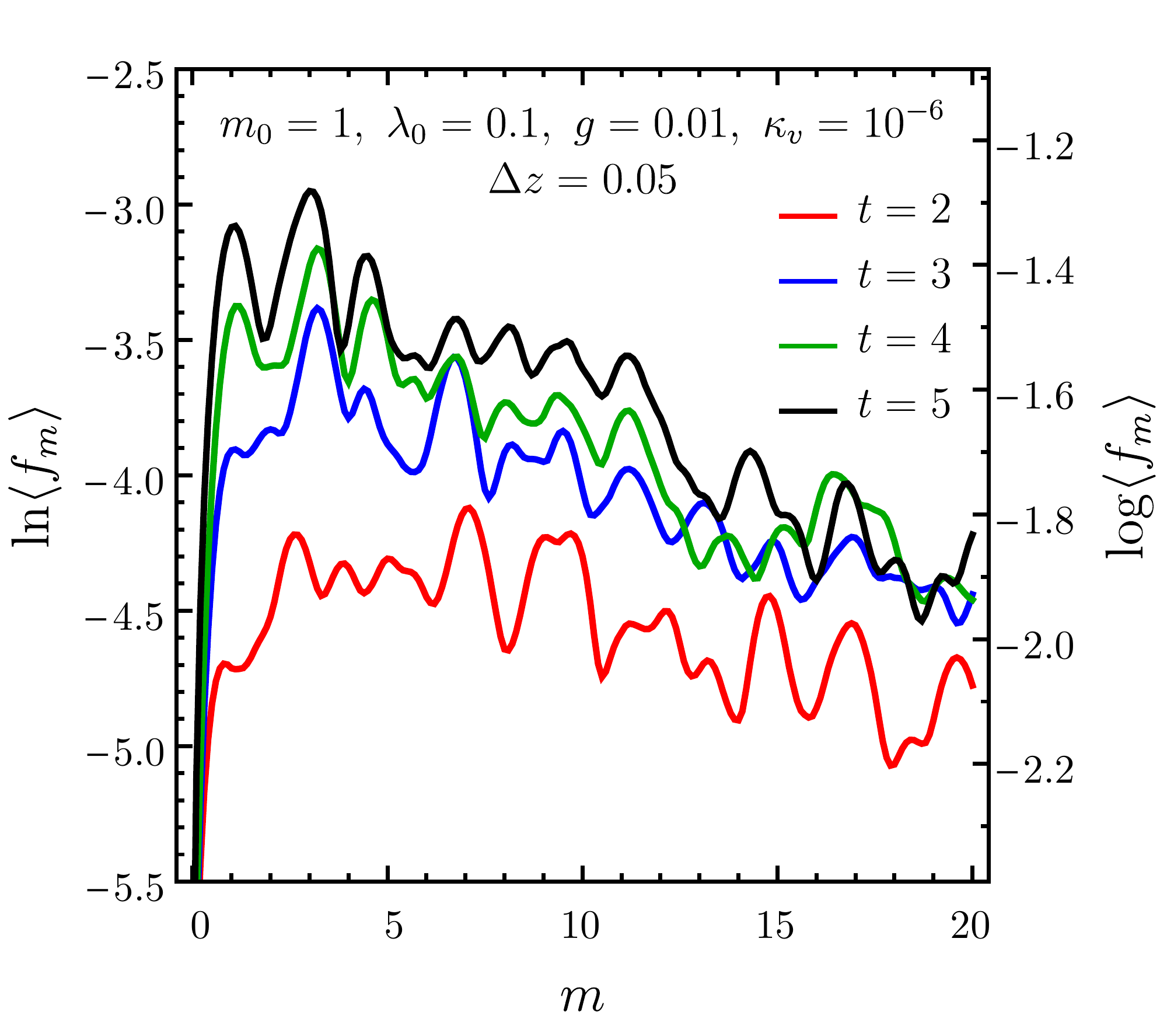}
    \caption{(\textbf{Left}) Growth of the density perturbation mode amplitude for the 2D simulation 
    shown in Fig. \ref{fig:2D-m1-g001}. The spatially averaged Fourier mode amplitude for a given 
    mode $m$ grows exponentially with time in the linear regime, such that $\langle f_m\rangle \propto e^{\eta_m t}$, 
    where $\eta_m$ is the growth rate of the mode. Different curves show the growth rate of the single wavelength 
    mode ($m=1$) averaged in the $\hat z$-direction over different mixing regions of size $\Delta z$ centered at 
    the middle of the hot unmagnetized layer. Also shown is the dependence of $\langle f_1\rangle$ on the magnitude of the initial 
    velocity perturbation $\kappa_v$, while keeping $\Delta z = 0.05$. The prediction of the linear theory 
    \textcolor{black}{(with arbitrary normalization)} from Eq.(\ref{eq:general-disp-rel}) for a mode with wavelength 
    $\lambda \approx \lambda_0$ is shown with dashed lines. (\textbf{Right}) Temporal evolution of the mode spectrum.}
    \label{fig:grate-m1-g001}
\end{figure*}

To study the linear and non-linear growth rates and structure of the KSI, we have conducted MHD 
simulations in 2D using the publicly available code \textit{Athena} \citep[v4.2;][]{Stone+08,BS11}. 
\textit{Athena} is a grid-based code that can solve the equations of relativistic MHD (RMHD) using 
Godunov methods. The problem under study is inherently relativistic with adiabatic index $\hat\gamma = 4/3$ 
in the relativistically hot plasma layer, and therefore, we used the special relativity module of 
\textit{Athena} with the HLLD Riemann solver, a third-order reconstruction of the primitive variables, 
and Van-Leer integrator.

The KSI is simulated in a box of size $(L_y,L_z) = (0.1,0.2)$, where $-0.05 \leq y \leq 0.05$ and 
$-0.1 \leq z \leq 0.1$. The current layer has (fluid-frame) width $\Delta = 0.01$ and the initial setup for 
the anti-aligned magnetic field case is shown in Fig. \ref{fig:sim-setup}, where the strength 
of the equilibrium magnetic field is set by the magnetization $\sigma$ of the cold plasma layer 
(here we set $c=1$ and use Lorentz-Heaviside units so that factors of $4\pi$ are ignored; see \citet{BS11} 
for equations of relativistic MHD in these units)
\begin{equation}
\sigma_0 \equiv\frac{w_{B,0}}{w_{m,0}}=\frac{b_0^2}{\rho_0+4p_g}\xrightarrow[\text{cold}]{}\frac{b_0^2}{\rho_0}~,
\end{equation}
where an adiabatic index of $\hat\gamma = 4/3$ has been assumed for a relativistically hot gas. 
The mass density in the cold magnetized layers is assumed to be $\rho_{0,c} = \rho_0 = 1$ with a density contrast 
$\rho_{0,h}=3\rho_{0,c}$ in the hot unmagnetized layer. At the initial moment, all layers are assumed to be in pressure 
equilibrium with homogeneous pressure $p_0 = \sigma_0/2$. The characteristic velocity of the system is the relativistic 
Alfv{\'e}n velocity
\begin{equation}
    v_A = \sqrt{\frac{\sigma_0}{1+\sigma_0}}=\sqrt{f_\sigma},
\end{equation}
where $0\leq f_\sigma \equiv \sigma_0/(1+\sigma_0)\leq 1$ is the fraction of the total energy in magnetic fields, 
and for which the crossing time $t_A = L_z/v_A$. Pressure homogeneity is maintained so long the condition given in
Eq.~(\ref{eq:delta_condition}) is met, with the magnitude of the effective gravity setting the scale of the simulation 
box $z\ll g^{-1}$. The code uses reflective boundary conditions in the $\hat z$-direction and periodic boundary conditions 
in the $\hat y$ direction. To initiate the instability, the equilibrium state is perturbed by introducing a velocity 
perturbation of the form
\begin{equation}
v_{1z}(y,z) = \frac{v_1}{2}\left[\sin\fracb{2\pi m_0 y}{L_y}\right]\left[1+\cos\fracb{2\pi z}{L_z}\right]~.
\end{equation}
where $m_0$ is the mode number in the $\hat y$-direction and the corresponding wavelength is $\lambda_0 = L_y/m_0$. 
This form of the perturbation ensures that no net momentum is imparted to the fluid elements in the simulation 
box. In addition, it is ensured that the perturbation velocity vanishes at the left and right boundary of the simulated 
region in order to suppress any spurious boundary effects. The perturbation amplitude is set to be a small fraction of the 
Alfv{\'e}n speed, such that $\kappa_v\equiv v_1/v_A\ll1$. 

To measure the growth rate of the instability in the linear 
stage, we follow the treatment by \citet{JNS95} and write the Fourier amplitude of the density perturbations as
\begin{equation}
f_m(z) = \frac{1}{L_y}\left\vert\int_{-L_y/2}^{L_y/2}\delta\rho(y,z)e^{-i2m\pi y/L_y}dy\right\vert
\end{equation}
where the perturbations are obtained by using the measure $\delta\rho(y,z) = [\rho(y,z)/\bar\rho(z)]-1$, which gives the amplitude 
of the density departure from the average density
\begin{equation}
\bar\rho(z) = \frac{1}{L_y}\int_{-L_y/2}^{L_y/2}\rho(y,z)dy~.
\end{equation}
The growth for a given mode $m$ is then obtained by averaging the mode amplitude over different length scales 
$\Delta z = z_2 - z_1$ around the perturbed hot layer,
\begin{equation}
\langle f_m\rangle = \frac{1}{z_2-z_1}\int_{z_1}^{z_2}f_m(z)dz~.
\end{equation}

In the following, temporal evolution of fluid quantities is shown using the simulation time. The time-step 
for the simulation depends on the resolution $\{N_y,N_z\} = \{512,1024\}$ and the CFL number $\mathfrak{C} = 0.5$, 
such that $\Delta t = \mathfrak{C}~{\rm Min}[L_z/N_z,L_y/N_y] = 0.977\times10^{-4}$. 


\begin{figure*}
    \centering
    \includegraphics[width=\textwidth]{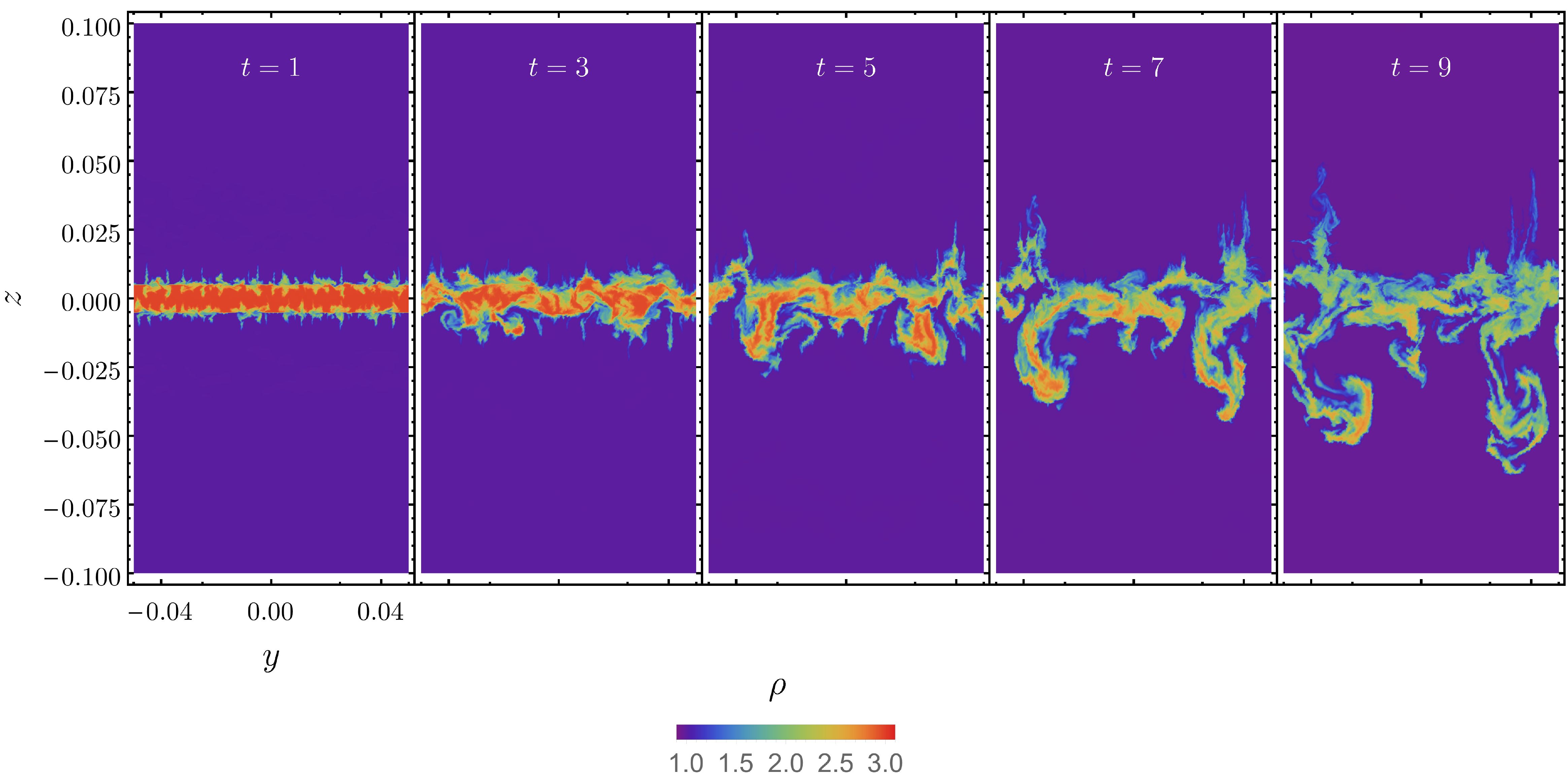}
    \caption{Development of the KSI for a higher value of $g = 0.1$, with the rest of the 
    parameters same as Fig.~\ref{fig:2D-m1-g001}.}
    \label{fig:2D-m1-g01}
\end{figure*}

\begin{figure*}
    \centering
    \includegraphics[width=0.48\textwidth]{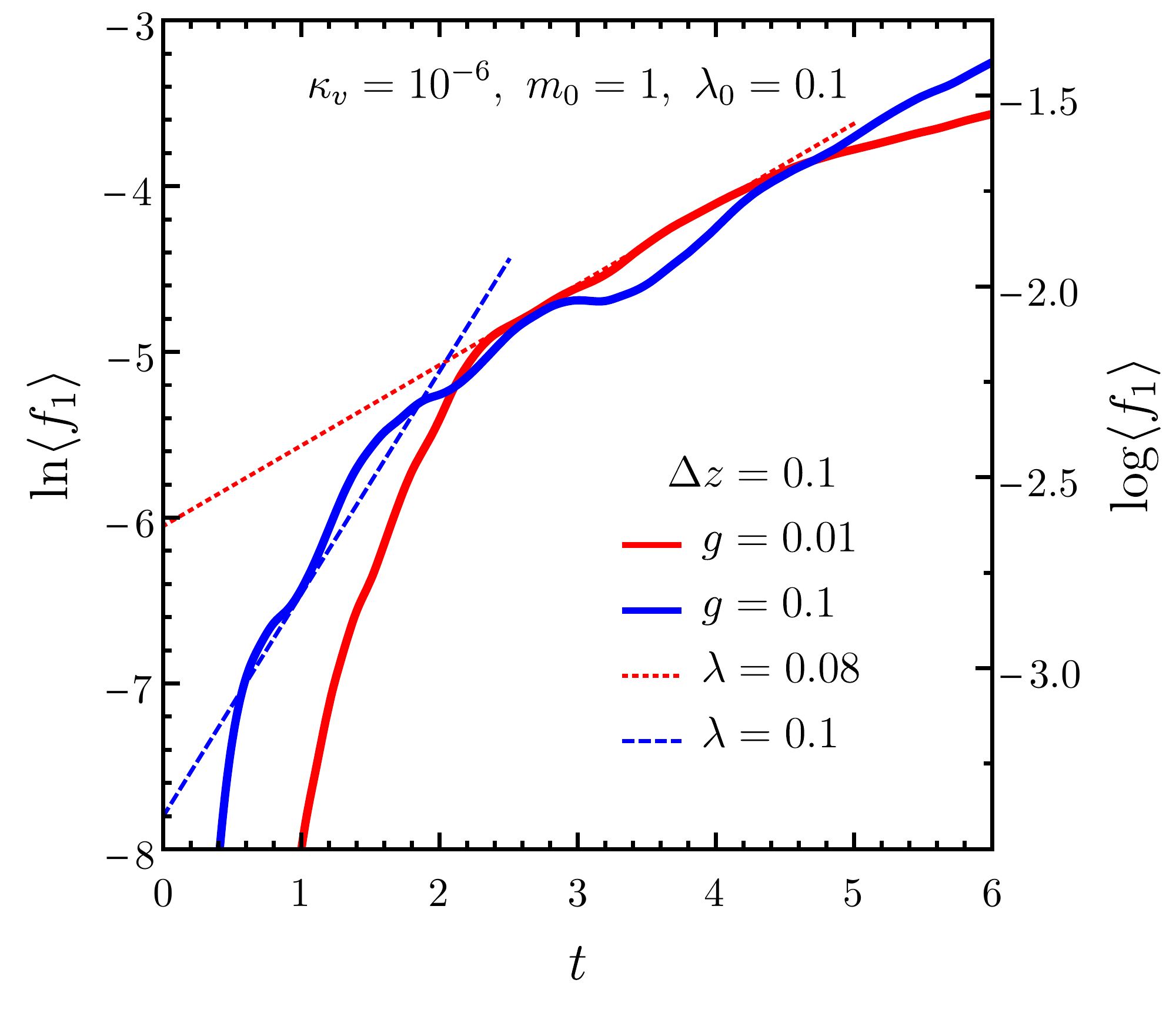}
    \includegraphics[width=0.48\textwidth]{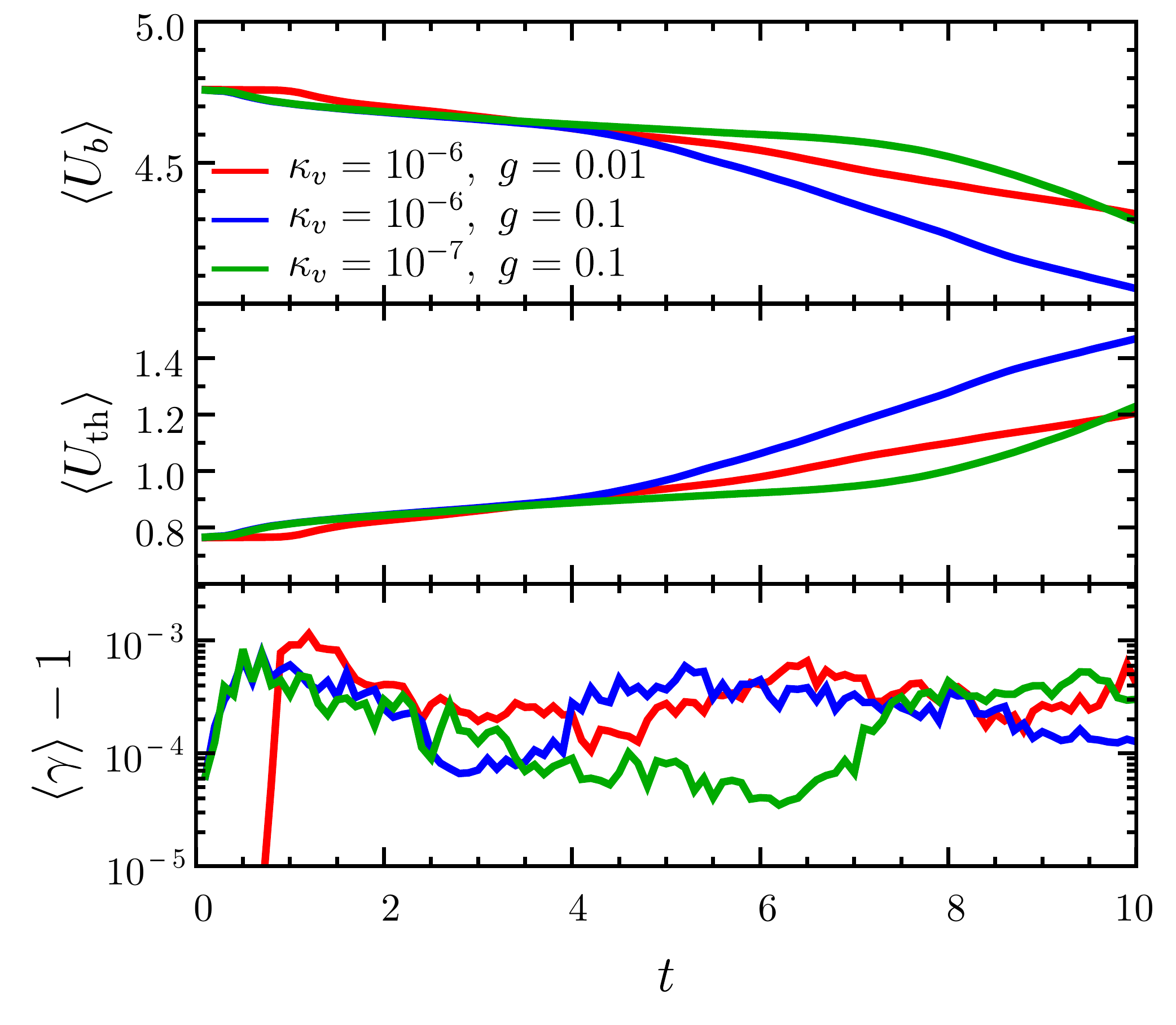}
    \caption{(\textbf{Left}) Comparison of growth rates obtained for the two cases with $m_0=1$ and $g=\{0.01,0.1\}$ to that expected from 
    linear theory (dashed \& dotted lines; \textcolor{black}{with arbitrary normalization}). (\textbf{Right}) 
    Temporal evolution of volume-averaged quantities for the single-wavelength ($m_0=1$) case shown in 
    Fig.~\ref{fig:2D-m1-g001}. Shown here are the magnetic field energy density (top), thermal energy 
    density (middle), and the Lorentz factor of fluid elements (bottom) for two different velocity perturbation 
    amplitudes and different effective gravity.}
    \label{fig:time-evol-m1-g001}
\end{figure*}

\begin{figure*}
    \centering
    \includegraphics[width=\textwidth]{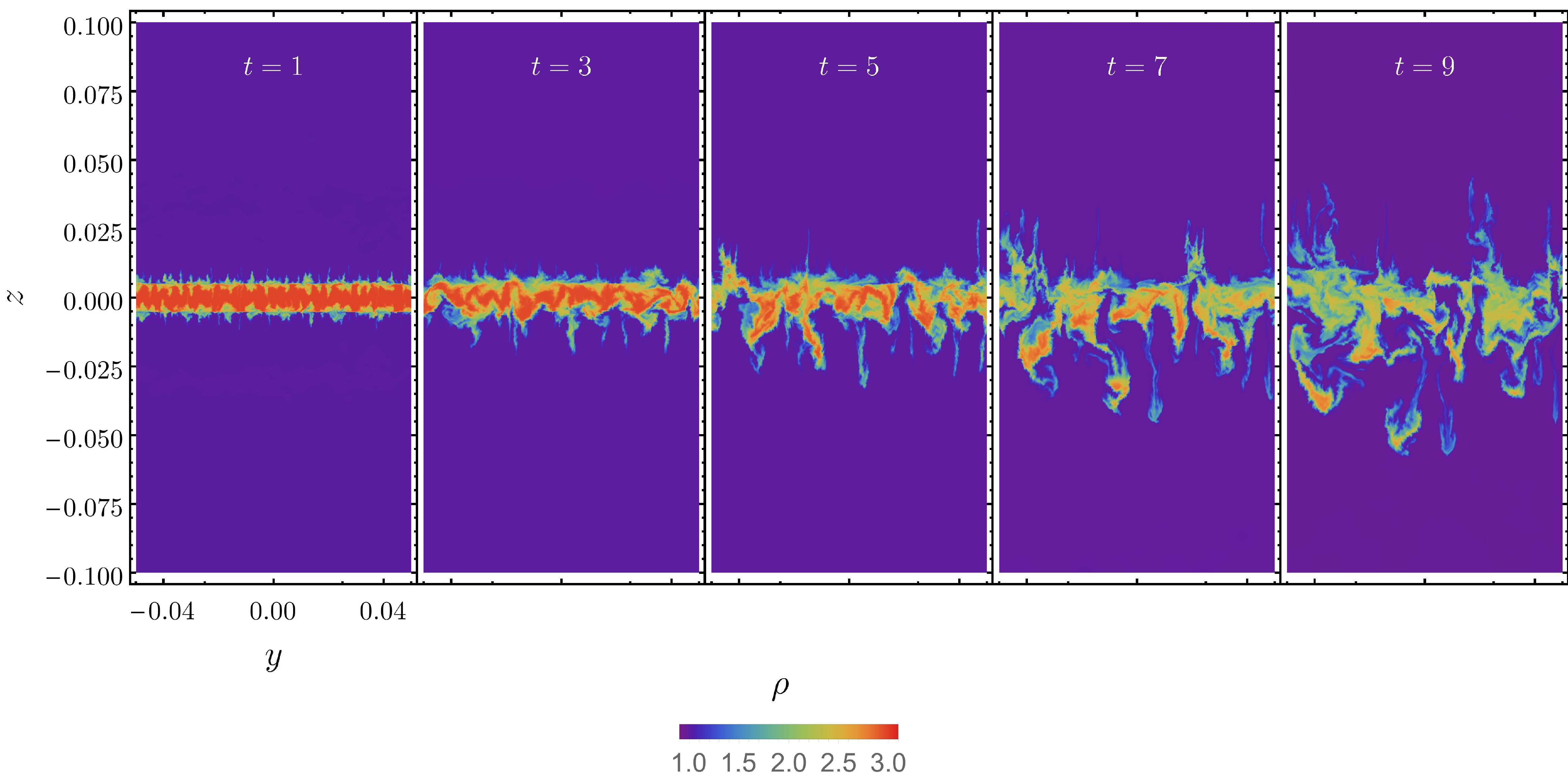}
    \caption{Development of the KSI from the linear to the non-linear stage for the high mode number 
    ($m_0 = 10$) initial velocity perturbation case, with $\kappa_v = v_1/v_A = 10^{-7}$ and $g = 0.1$. The 
    other parameters are the same as in Fig.~\ref{fig:2D-m1-g001}.}
    \label{fig:2D-m10-g01}
\end{figure*}

\begin{figure*}
    \centering
    \includegraphics[width=0.3\textwidth]{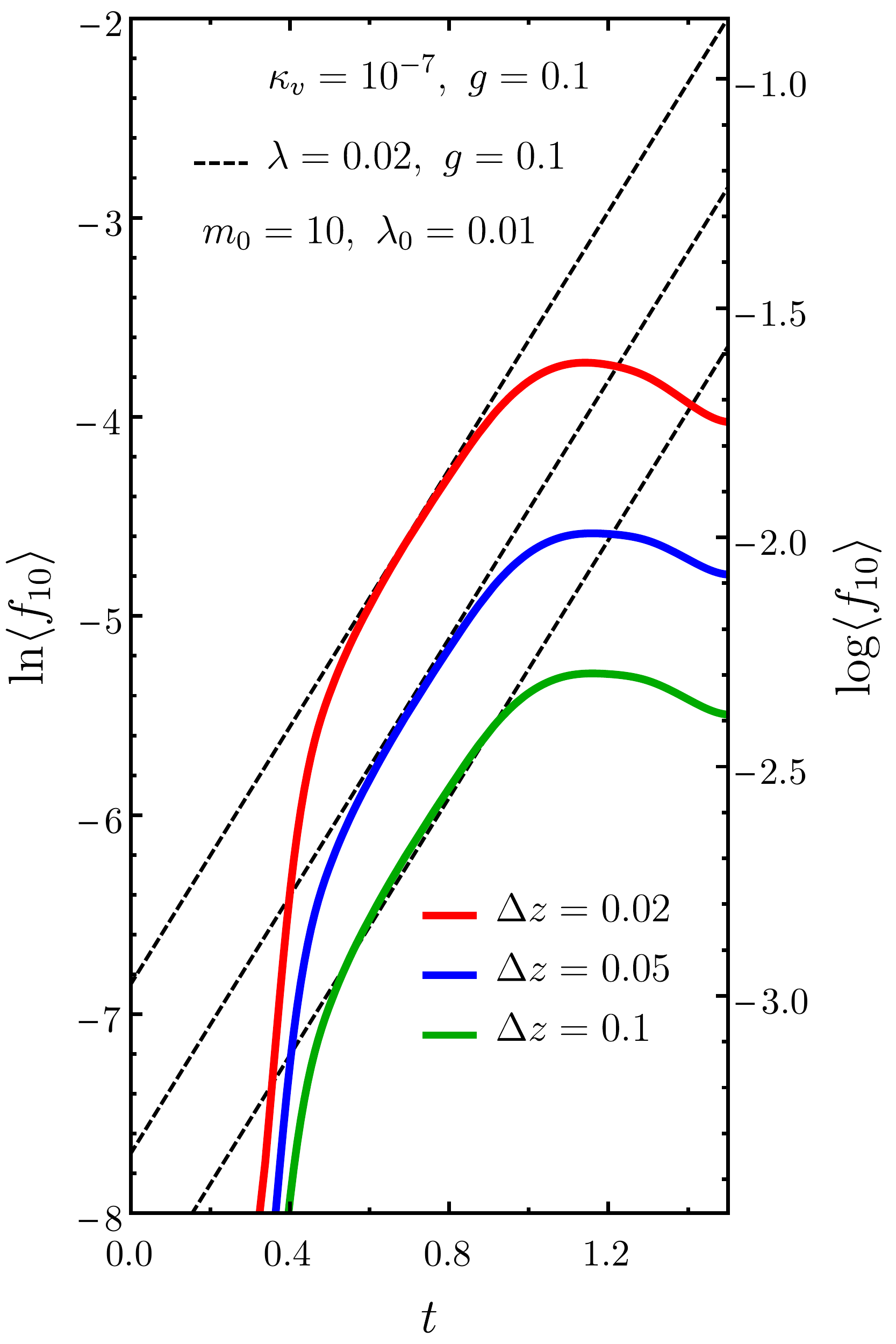}\hspace{1cm}
    \includegraphics[width=0.48\textwidth]{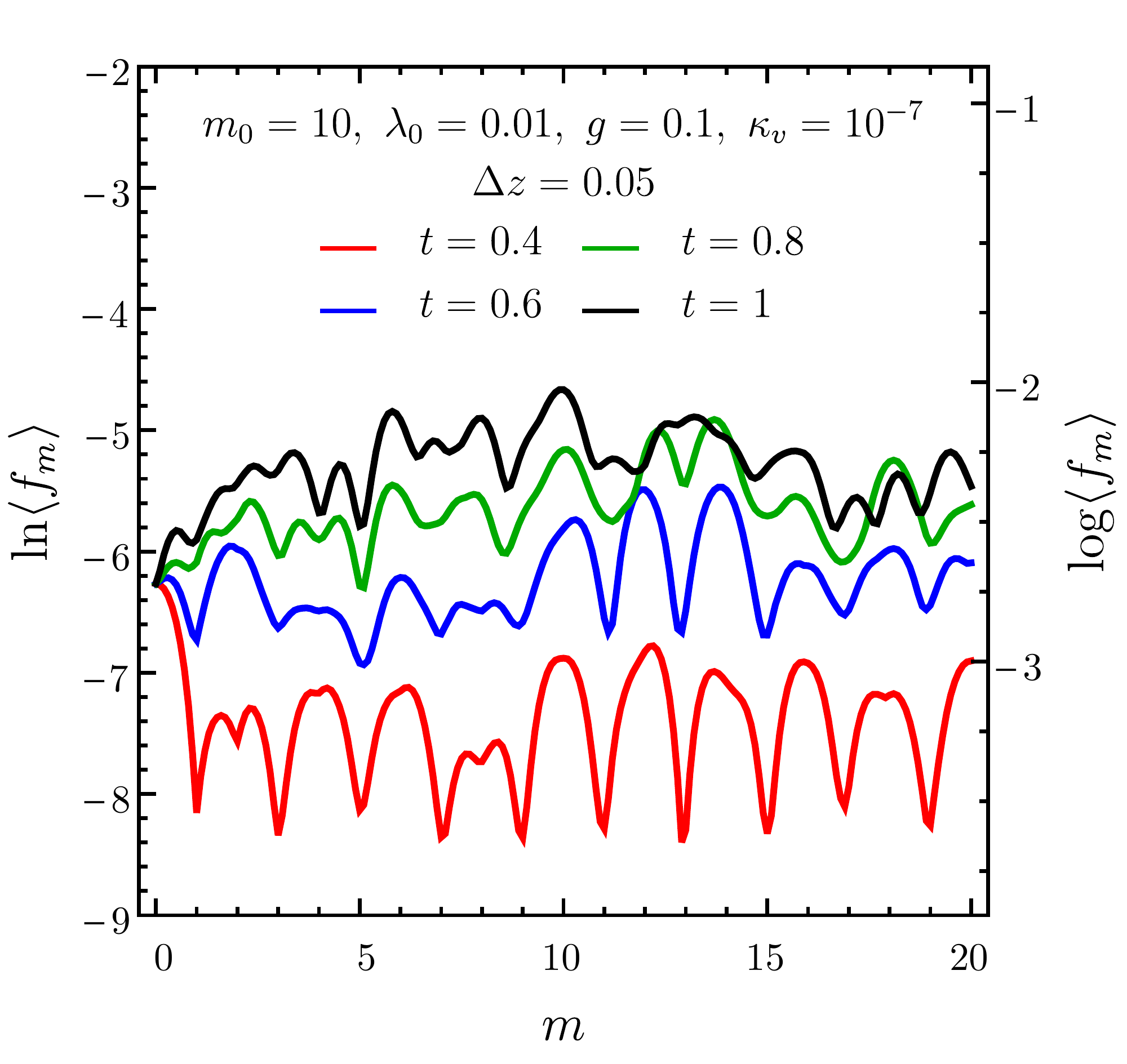}
    \caption{\textbf{(Left)} Growth of the density perturbation mode amplitude for the 2D simulation 
    shown in Fig.~\ref{fig:2D-m10-g01}. The three curves correspond to the Fourier amplitude of the 
    mode $m = 10$ averaged over different mixing regions of size $\Delta z$ centered at the middle 
    of the hot unmagnetized layer. The prediction of the linear theory (\textcolor{black}{with arbitrary normalization}) 
    from Eq.~(\ref{eq:general-disp-rel}) for 
    a mode with wavelength $\lambda \approx\lambda_0$ is shown with dashed lines. (\textbf{Right}) Temporal evolution 
    of the mode spectrum in the linear stage for the simulation shown in Fig.~\ref{fig:2D-m1-g01}. 
    }
    \label{fig:grate-m10-g01}
\end{figure*}

\begin{figure*}
    \centering
    \includegraphics[width=0.95\textwidth]{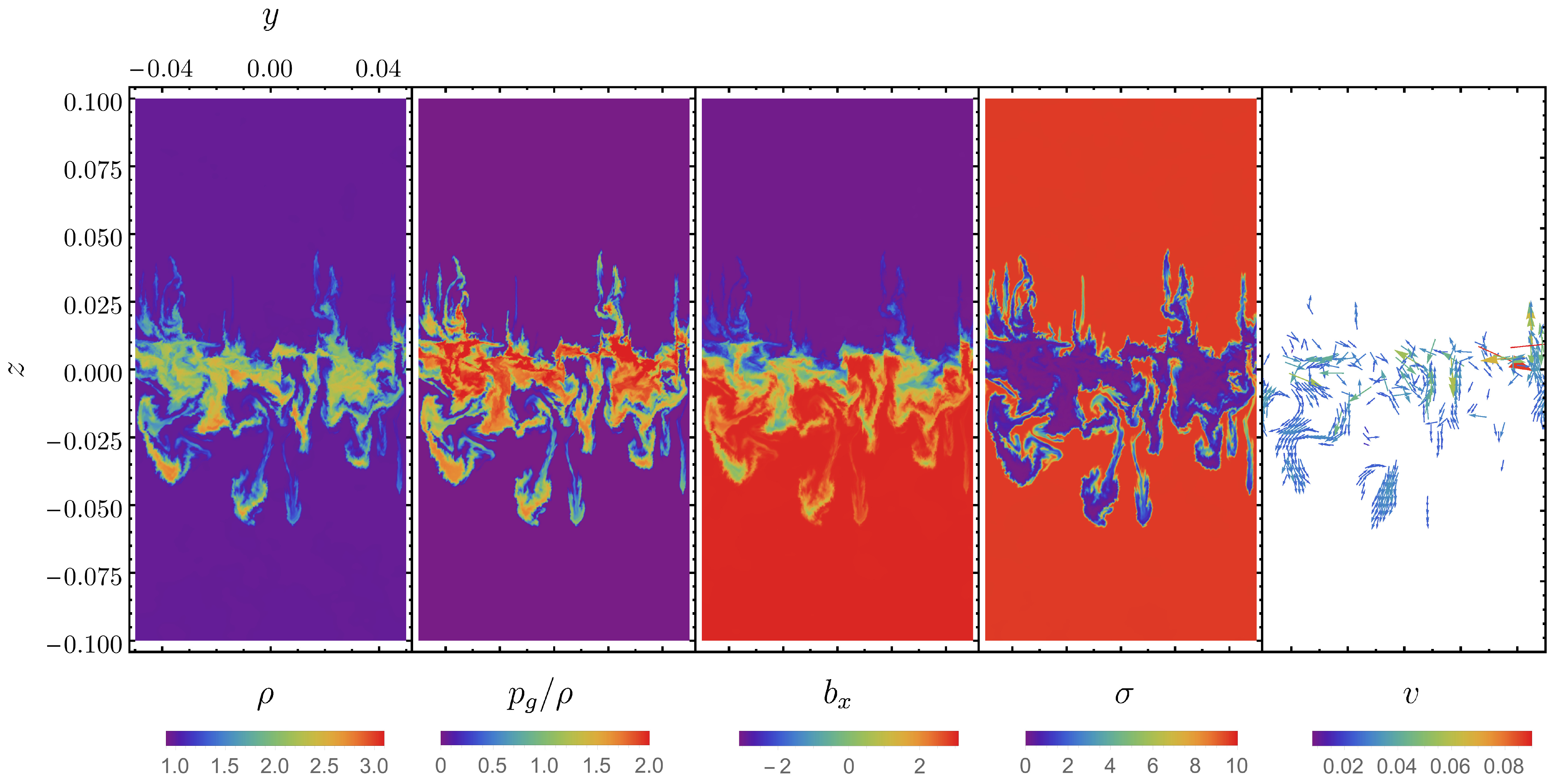}
    \caption{State of various quantities for the high mode number ($m_0 = 10$) perturbation 
    with $g=0.1$ at $t=9$. Shown here are the gas density ($\rho$), 
    ratio of gas pressure and density ($p_g/\rho$), the $\hat x$-component of the magnetic field ($b_x$), 
    magnetization ($\sigma$), and gas velocity ($\mathbf{v}$).}
    \label{fig:ks2D-multi-par-compare}
\end{figure*}

\begin{figure*}
     \centering
    \includegraphics[width=0.48\textwidth]{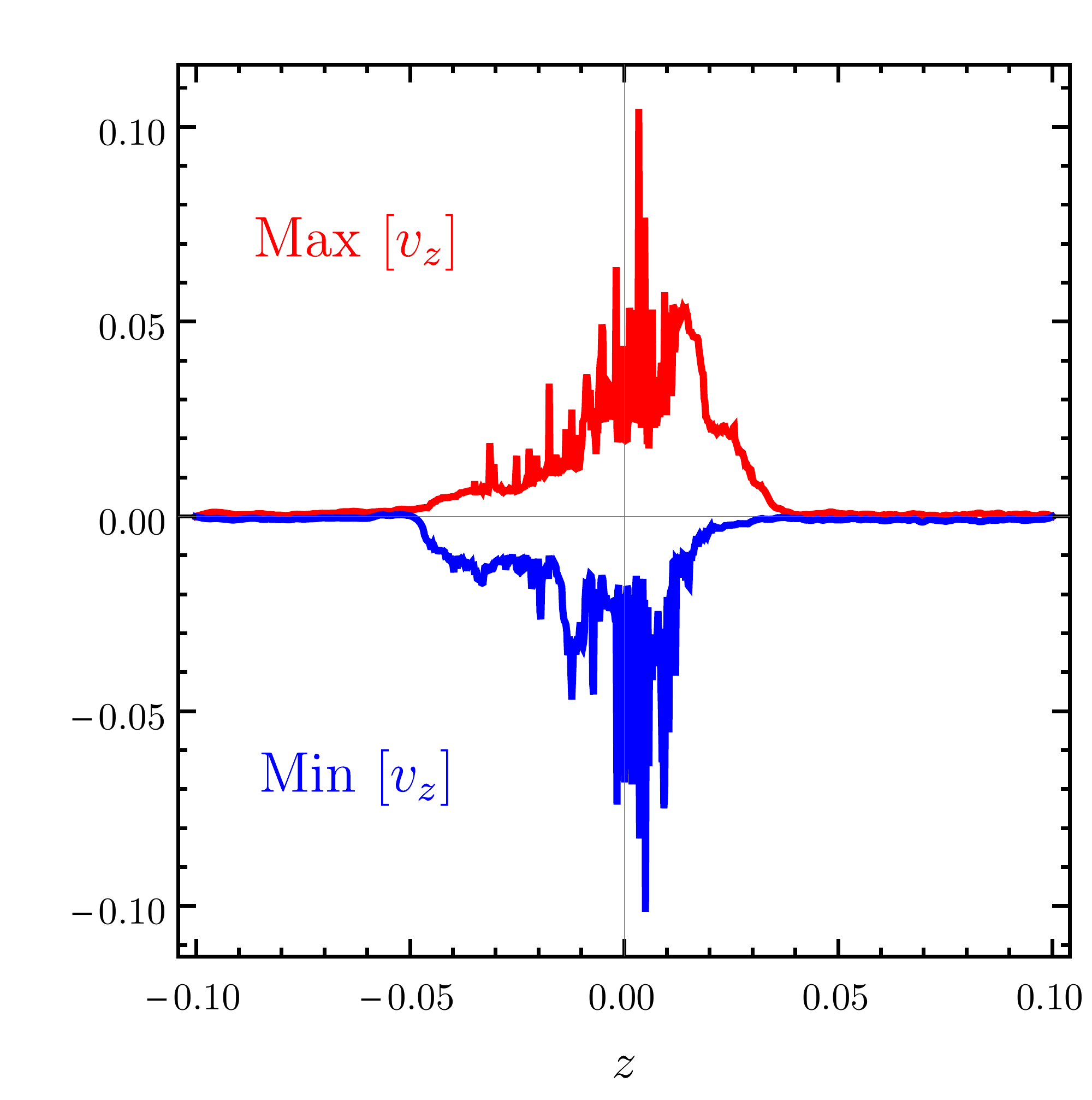}\hspace{.2cm}
    \includegraphics[width=0.48\textwidth]{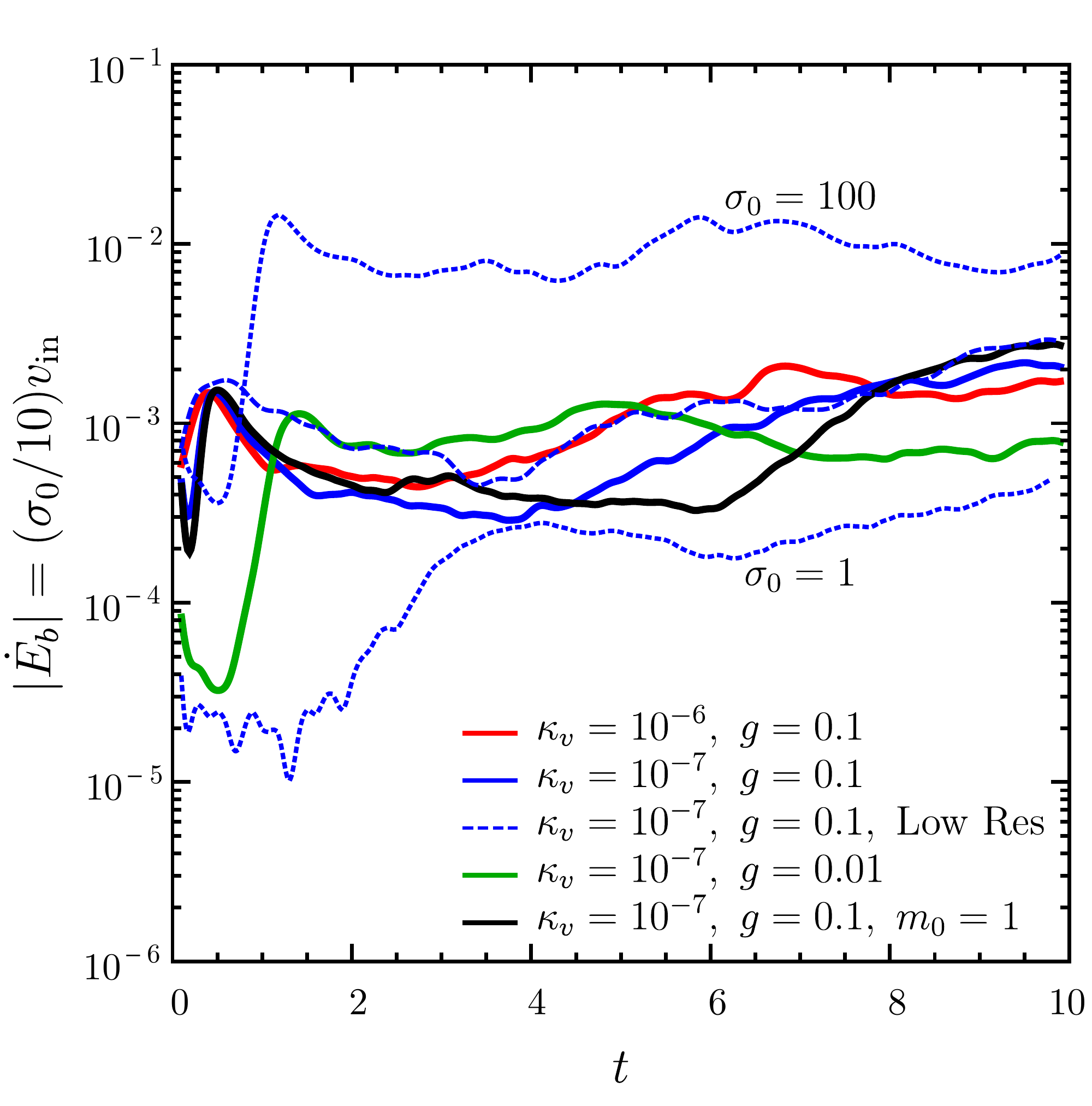}
    \caption{\textbf{(Left)} Peak velocity in the $\hat z$-direction at a given $z$-coordinate and across all $y$ values, 
    for the simulation shown in Fig.~\ref{fig:ks2D-multi-par-compare} with high mode number ($m_0=10$) perturbation and $g=0.1$. 
    In the highly mixed region, near the middle of the simulation box where the current layer was initialized,
    fluid velocities reach $\vert v_z\vert\sim 0.1$, where the initial velocity perturbation amplitude was 
    $10^{-7}v_A$. The peak $v_z$ shows no clear dependence on $\kappa_v$, $g$, or $\sigma_0$. 
    \textbf{(Right)} Magnetic energy dissipation rate due to reconnection in the simulated volume 
    ($\vert \dot E_b\vert \approx L_yL_zb\vert\dot b\vert$) for $m_0=10$ perturbation. The result from a 
    low-resolution [$256\times512$] simulation is also shown.}
    \label{fig:peak-vz-diss-rate}
\end{figure*}

\section{Simulation Results}\label{sec:results}
In 2D the instability mode can either be transverse or parallel to the equilibrium magnetic field. 
Here we present results for the transverse case, which as expected evolves similarly to the hydrodynamic 
case since the magnetic field doesn't play any role in stabilizing the instability. In the parallel case, 
unstable perturbations can be stabilized by magnetic field tension.

\subsection{Large Wavelength ($k\Delta\ll1$) Mode}
It is instructive to first study the much simpler single wavelength mode, $m_0=1$, before undertaking more 
complicated scenarios. Fig.~\ref{fig:2D-m1-g001} demonstrates the development of the KSI for $m_0=1$ by showing snapshots of the plasma rest mass density $\rho$ at different simulation times. The anti-aligned 
equilibrium magnetic field lines go into (out of) the page in the region above (below) the hot layer. 
The wavelength of the initial velocity 
perturbation is $\lambda_0 = L_y$ and we choose a small enough $\kappa_v$ so that the instability in driven only by the effective gravity and not by the initial conditions. To that end, we have carried out 
simulations to establish the upper limit on $\kappa_v<10^{-5}$ below which the current layer remains stable 
for $g=0$, such that the initial perturbations don't grow over time. On 
the other hand, a high $\kappa_v>10^{-5}$ disrupts the hot layer, regardless of the magnitude of $g$, and 
drives the mixing of the two magnetized fluids, however, the instability in this case is artificial and not 
driven by the effective gravity.

The mode amplitude, averaged over the mixing region of extent $\Delta z$, should grow exponentially with time 
in the linear stage, such that $\ln\langle f_m\rangle\propto \eta_m t$. We plot this quantity for different sizes 
of the mixing region, $\Delta z$, in the left-panel of Fig.~\ref{fig:grate-m1-g001} and compare it with predictions from the linear 
theory, \textcolor{black}{which are shown with arbitrary normalization since only their slope is relevant here}. The different stages of the instability shares many similarities with the RTI (compare with e.g. Fig. 6 of \citet{JNS95}) 
and can therefore be understood in a similar manner.

The instability proceeds in three separate stages and its 
growth rate at any given time can be inferred from the slope of the curves in Fig.~\ref{fig:grate-m1-g001}. (1) \textit{Super-linear 
stage}: This stage simply reflects the response of the current layer to the initial condition where the interface 
is disturbed by the initial velocity perturbation. This excites several small wavelength ($m>1$) modes at early times as can be seen in the mode 
spectrum shown in the right-panel of Fig.~\ref{fig:grate-m1-g001}. At this stage, the spatially averaged mode amplitude 
$\langle f_m\rangle$ grows super-exponentially. (2) \textit{Linear stage}: Very quickly ($\sim 20$ light-crossing times of the width of the 
box), the instability enters the linear stage where the growth rate is determined by the dominant $m\sim m_0=1$ mode and the magnitude of 
the effective gravitational acceleration; the power in modes much larger than $m_0$ is relatively small in this stage. Due 
to significant mixing of modes higher than $m_0$, the resultant growth rate corresponds to a mode with $\lambda = 0.08 \lesssim \lambda_0=0.1$. 
The linear stage commences at the same time in all averaged regions of size $\Delta z$, which is expected since 
$\Delta z$ is centered at the current layer. The same behavior is seen for the time at which the instability enters 
the non-linear stage. \textcolor{black}{In addition, since the growth rate of the instability in the linear regime only 
depends on the magnitude of the effective gravity and initial wavelength of the mode, it is not expected to vary with 
$\Delta z$. The growth rate curves shown in Fig.~\ref{fig:grate-m1-g001} show excellent agreement with this expectation.} 
(3) \textit{Sub-linear stage}: Eventually, the non-linear stage of the instability sets in and the growth rate of the 
instability slows down. This is marked by a plateau in the Fourier mode amplitude with time. 

It is important to stress here that the ``linear" stage -- stage (2) above -- that is used for comparison, actually corresponds to $\ln\langle f_m\rangle\propto \eta_m t$, i.e. a linear growth of the logarithm of the perturbation amplitude $\langle f_m\rangle$, and therefore to an exponential growth of $\langle f_m\rangle$ itself.

The growth rate of the instability scales as $\eta \propto g^{1/2}$, and therefore a higher 
effective gravity $g$ should lead to a more evolved state at a given simulation time $t$. This can indeed be clearly seen in Fig.~\ref{fig:2D-m1-g01}, which shows the evolution of the KSI for a higher effective gravity, $g = 0.1$ (e.g. by the deeper penetration of the density fingers 
below the hot current layer). Furthermore, the asymmetry between the upper and lower regions is now much clearer. It is 
interesting to note that the hot plasma dripping from the current layer does not simply move downwards in an approximately 
straight line, but instead starts to curl upwards. This effect is investigated in more detail in \S~\ref{sec:buoyancy} where 
we discuss the effects of buoyancy and secondary plasma instabilities on the dripping blobs.

In the left-panel of Fig.~\ref{fig:time-evol-m1-g001}, we compare the growth rate of the mode amplitudes for the two 
cases with $g=\{0.01,0.1\}$, where the higher effective gravity causes the instability to reach the linear stage earlier in 
time as compared to the lower $g$ case. 
In the non-linear stage, a higher $g$ causes the unmagnetized over-dense regions to \textit{drip} faster into 
the lower cold magnetized region \citep{L10}. This displacement of large volumes of the hot layer downwards is reciprocated by 
the upward movement of bulk plasma from the lower region. The meeting of the rising bulk plasma blobs with thinner regions of 
the hot layer leads to the launching of thin plumes upwards (like when a stone is thrown into water).

The dripping of the unmagnetized fluid brings the two magnetized fluids of the striped wind into contact that undergo resistive dissipation of the entrained magnetic field. This has the effect of destroying magnetic energy and consequently heating up the fluid. As a result, the thermal energy should increase due to increase in the amount of hot fluid. The 
temporal evolution of volume averaged magnetic energy density $\langle U_b\rangle$, thermal energy density 
$\langle U_{\rm th} \rangle$, and 
Lorentz factor of fluid elements $\langle\gamma\rangle-1$ is shown in the right panel of Fig.~\ref{fig:time-evol-m1-g001}. 
Since the bulk motion of the fluid elements remains non-relativistic, most of the dissipated energy goes into the internal energy 
of the relativistic gas rather than the kinetic energy. Although the growth rate of the mode amplitude did not show any 
dependence on $\kappa_v$, the temporal evolution of the magnetic energy, and correspondingly the thermal pressure, indeed does. 
A larger $\kappa_v$ yields a faster magnetic energy decay rate in the earlier part of the non-linear phase; the two rates are 
similar in the later part. The former result is also true for the case with higher effective gravity.

\subsection{Wavelength Comparable to the Size of the Current Sheet ($k\Delta\approx1$)}
Higher mode number perturbations ($m_0 > 1$) with wavelength comparable to the size of the current layer are 
most interesting as they are expected to yield vigorous disruption of the current layer. To achieve that 
we introduce perturbations with mode number $m_0 = 10$ or equivalently with wavelength 
$\lambda_0 = \Delta = 0.01$. The development of the instability in this case for $g=0.1$ is shown in Fig.~\ref{fig:2D-m10-g01}. 
In comparison to the single-wavelength case, the instability clearly shows more structure, which simply reflects the 
smaller wavelength of the seed perturbation. In both cases, the density fingers eventually penetrate to 
approximately similar depths at similar simulation times. This is expected since the effective gravity is the same in both cases.

In the linear stage, the instability grows at the rate for a mode with $\lambda\sim\lambda_0$, as shown in the left-panel 
of Fig.~\ref{fig:grate-m10-g01}, where excitation of various modes both with $m<m_0$ as well as $m>m_0$ can be seen 
in the right-panel. The mode spectrum is more complex in this case and lacks the clear suppression of power for modes 
away from $m_0$, as was seen in the single wavelength case.

To glean further information regarding the dynamical evolution of the instability, in Fig.~\ref{fig:ks2D-multi-par-compare} we show 
the state of the thermal pressure to rest mass density ratio $p_g/\rho$, magnetic field orthogonal to the page $b_x$, magnetization $\sigma$, and the 
fluid velocity $\mathbf v$ in the $y$-$z$ plane, along with the density $\rho$ for comparison. These quantities are shown at 
time $t=9$ where their state serves as a proxy to the amount of mixing that has occurred between the magnetized and unmagnetized fluids as well as the two magnetized fluids with oppositely oriented magnetic field lines. The 
degree of mixing between the magnetized and unmagnetized fluids, as it appears in the plot showing $\sigma$, is a good 
indicator of the amount of magnetic diffusivity that artificially dissipates magnetic energy. On the other hand, 
the level of mixing between the two magnetized fluids, as can be seen most clearly in the plot showing $b_x$, is 
what determines the rate of magnetic energy dissipation. 
We stop the simulation before the 
density fingers reach the bottom of the simulation box to avoid the final solution from being affected by the boundary conditions. 
Since there is no large scale directed flow towards the hot layer, the mixing is purely determined by the action of the effective 
gravity. The unmagnetized hot fluid has to drip out before the two magnetized regions can come into contact and undergo magnetic 
reconnection. 

\section{Reconnection Rate}\label{sec:Rec-rate}
Most magnetic reconnection models feature an ordered bulk flow towards the current layer, in which case the velocity of this bulk flow, $v_{\rm in}$, just upstream of the current layer provides a good measure of the reconnection rate.
However, this is a good measure of 
the reconnection rate only if the flow towards the current layer is ordered on large scales. In the simulations shown in this work, we find that the fluid motions in the mixing region are highly turbulent and lack order on larger scales. 
In the left-panel of Fig.~\ref{fig:peak-vz-diss-rate}, we plot the peak velocity in the $\hat z$-direction, which is orthogonal to the direction 
of the equilibrium magnetic field, $\hat{z} \perp \mathbf{b_0}$. The peak velocity of $\vert v_z\vert \sim 0.1$ is 
reached near the middle of the simulation box where the current layer was initialized and where the two magnetized 
regions undergo the maximum amount of mixing and reconnection. Near the top and bottom of the simulation box, where no reconnection is expected, the fluid velocity remains much smaller. 
Furthermore, practically the same level of peak velocity is attained, regardless of $\kappa_v$, $g$, and $\sigma_0$, in all the simulations 
that we performed. This suggests that $\vert v_z\vert \sim 0.1$ is the maximum attainable turbulent velocity. 
Nevertheless, it appears that in our physical setup $\vert v_z\vert$ does not serve as a good measure of the actual
reconnection rate.

An alternative measure of the reconnection rate that works better in our case may be obtained by directly calculating from the simulation results the rate at which the magnetic field energy is dissipated. 
The temporal evolution of the dissipated power, 
\begin{equation}
-\dot E_b = -\frac{d}{dt}\int_{-L_z/2}^{L_z/2}dz\int_{-L_y/2}^{L_y/2}dy~\frac{1}{2}b^2(y,z)~,
\end{equation}
for both $m_0=1$ and $m_0=10$ cases is shown in the right-panel of Fig.~\ref{fig:peak-vz-diss-rate}, 
including results from a low-resolution (Low Res: $256\times512$ ) simulation. 
In all cases, there is a surge in the dissipation rate at very early times followed by saturation of the 
rate, which happens very quickly. The simulation is stopped at $t = 10$ at which point the density 
fingers almost reach the bottom of the simulation box. The magnetic field energy dissipation rate, 
after it has saturated, allows a straightforward determination of the reconnection rate,
\begin{equation}\label{eq:v_in}
    \vert \dot E_b\vert = 2v_{\rm in}L_y\frac{b_0^2}{2} = \sigma_0\rho_{0,c}L_yv_{\rm in}~,
\end{equation}
which yields $\vert \dot E_b\vert = v_{\rm in}$ for $\sigma_0 = 10$, $\rho_{0,c}=1$ and $L_y = 0.1$. We stress that here $v_{\rm in}$ represents the velocity of an ordered bulk inflow that would produce the same magnetic energy dissipation rate as the one that is produced in our simulations where the central mixing region is highly turbulent with no clear bulk flow. In this sense it serves as an effective bulk velocity, which is useful mainly for the purpose of comparison with the results of other magnetic reconnection models.

\begin{figure}
    \centering
    \includegraphics[width=0.4\textwidth]{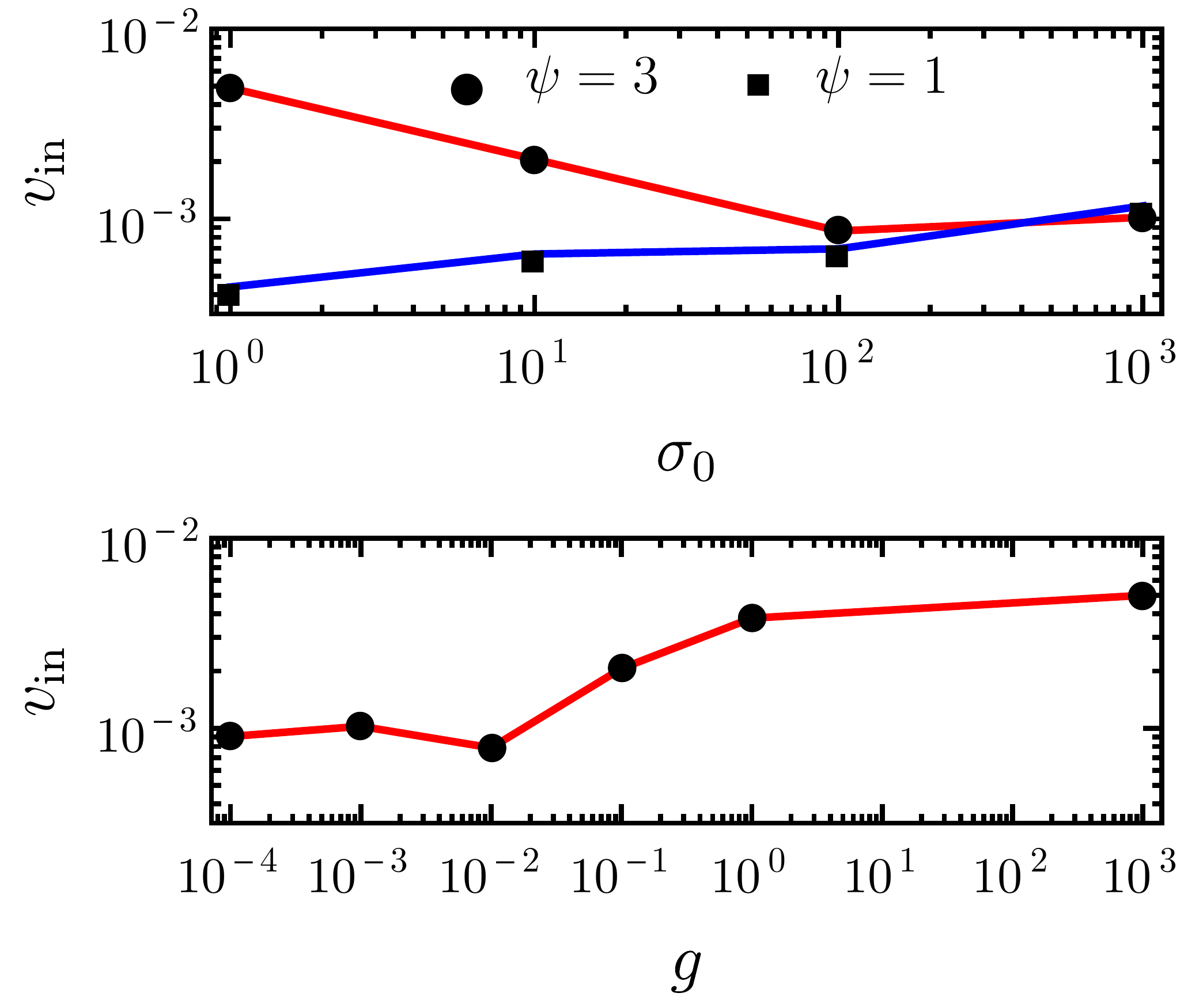}
    \caption{Reconnection rate as a function of the effective gravity $g$ and fluid magnetization $\sigma_0$. 
    The rate of reconnection is defined using an effective upstream velocity $v_{\rm in}$ that is inferred 
    from the rate of dissipation of magnetic field energy density. In the upper panel $g=0.1$ and $\sigma_0=10$ 
    in the lower panel; $\kappa_v = 10^{-7}$ for both panels. The dissipation rate for two values of the 
    density ratio $\psi\equiv\rho_{h,0}/\rho_{c,0}=\{1,3\}$ is shown as a function of $\sigma_0$ in the 
    top-panel.}
    \label{fig:vin-g-sigma}
\end{figure}

The dependence of $v_{\rm in}$ on $\sigma_0$ and $g$ is shown in Fig.~\ref{fig:vin-g-sigma}, which indicates that 
as $\sigma_0$ increases the efficiency of the reconnection process aided by the turbulent mixing declines. This 
can be understood as follows. The growth rate of the instability depends on the enthalpy density contrast between the 
hot and cold regions, such that $\eta\sim\sqrt{\Lambda gk}$ where $\Lambda=(w_{h,0}-w_{c,0})/(w_{h,0}+w_{c,0})$, and 
$w_{h,0}$ and $w_{c,0}$ are respectively the initial enthalpy densities of the hot and cold regions. 
For $\rho_{h,0}=\psi\rho_{c,0}=\psi$ and the initial pressure $p_0=b_0^2/2=\sigma_0\rho_{c,0}/2=\sigma_0/2$, 
we find 
\begin{equation}
\Lambda = \frac{1}{3}\left(1+\frac{2(\psi-2)}{1+\psi+3\sigma_0}\right)\xrightarrow[\sigma_0\to\infty]{}\frac{1}{3}~,
\end{equation}
where $\Lambda < 1/3$ for $\psi<2$, and $\Lambda > 1/3$ for $\psi>2$. 
Therefore, when $\sigma_0\sim\;$a few the magnitude of $\psi$ around the critical value of $\psi_c=2$ causes 
the growth rate of perturbations to diverge, as shown in the top-panel of Fig.~\ref{fig:vin-g-sigma}, which 
leads to a higher or lower $v_{\rm in}$; in the high-$\sigma_0$ limit $v_{\rm in}$ saturates regardless of the 
magnitude of $\psi$.

While in the high-$\sigma_0$ limit $v_{\rm in}\to10^{-3}$ for $g=0.1$, increasing the effective gravity shows a promising 
increase in the reconnection rate. However, the maximum value of $g$ in the simulation is limited by the dynamical time $t_{\rm dyn} = g^{-1} =10 - 100$ for $g = 10^{-1} - 10^{-2}$. 
This timescale should be longer than the free-fall time 
of a given fluid element where $t_{\rm ff}=\sqrt{2\delta z/g}=\sqrt{2}$ for $g=0.1$ and $\delta z=L_z/2=0.1$, the half length of the 
simulation box, i.e. one requires $t_{\rm ff}/t_{\rm dyn} = \sqrt{L_z/L_{\rm dyn}}<1$, which corresponds to $g<1/L_z = 5$ for $L_z=0.2$. The same result can be obtain from another consideration, that as $g$ 
increases the length scale over which the total pressure is homogeneous also shrinks, which necessitates the 
need for a pressure gradient. The initial pressure was assumed to be uniform in all simulations since the vertical 
length of the simulation box $L_z \ll L_{\rm dyn}=g^{-1}$.

The reconnection rate at $g\gg1$ can only be probed at scales much smaller than used thus far in all the 
simulations. In order to do that, we ran additional simulations with a box size that was smaller by a factor 
of $10^{-3}$ for $g=10^3$, $\sigma_0=10$, and $\kappa_v=10^{-7}$. The resulting $v_{\rm in}$ is shown as the last 
point in the bottom-panel of Fig.~\ref{fig:vin-g-sigma} that clearly shows the saturation of the reconnection 
rate at $v_{\rm in}\approx5\times10^{-3}$; this rate is expected to be slightly lower for $\sigma_0\gg1$.
\begin{figure*}
    \includegraphics[width=0.95\textwidth]{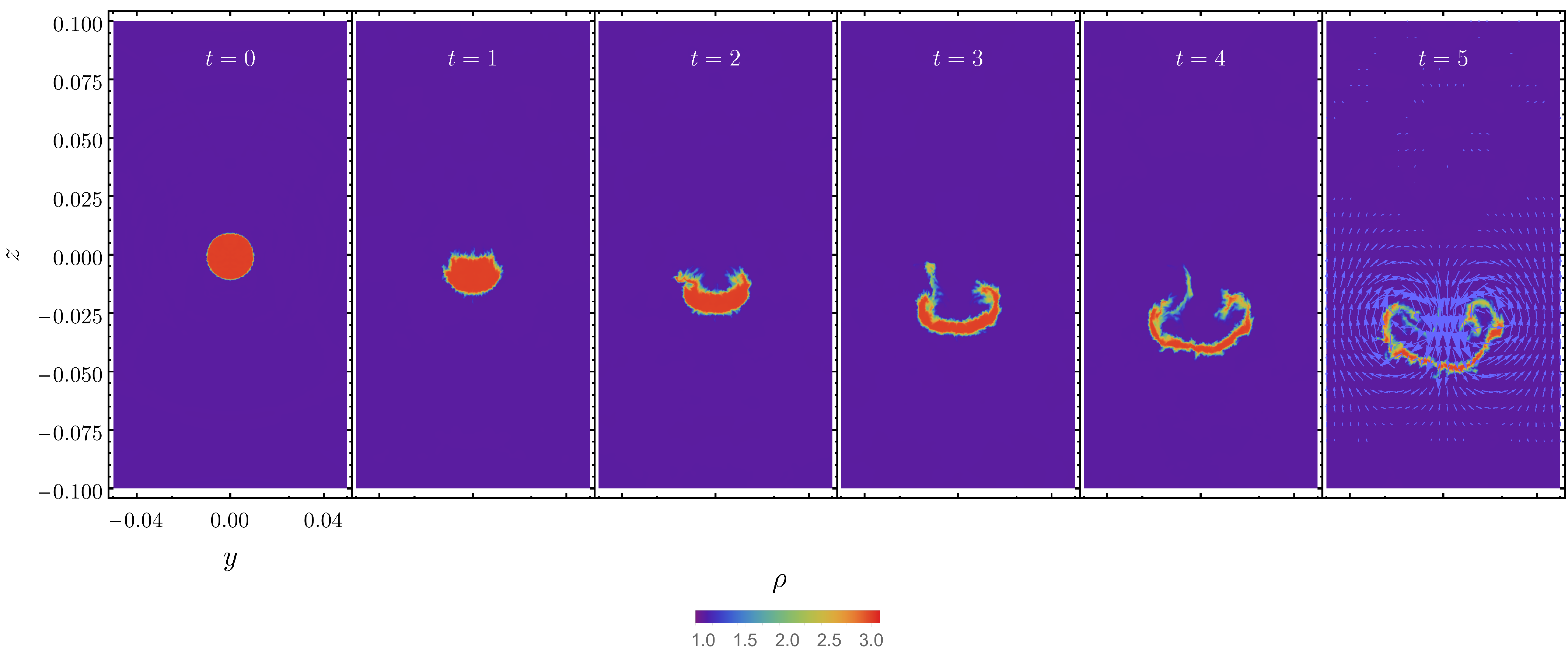}
    \caption{Density snapshots of an un-magnetized blob with initial downward velocity $v_b = 0.01$ in a magnetized fluid. 
    The uniformly magnetized fluid, with magnetic field pointing out of the page, is cold and has magnetization $\sigma=10$. 
    Gravity points downward with $g=0.1$. The blob is flattened due to the ram pressure of the ambient fluid. Onset of the Kelvin-Helmholtz instability leads to the development of symmetrical 
    vortices as shown in the last snapshot, which provide an effective drag force that approximately balances the effective gravity accounting for the buoyancy force. }
    \label{fig:blob}
\end{figure*}

\section{The Role of Buoyancy, Kelvin-Helmholtz Instability and Vorticity}\label{sec:buoyancy}
The KSI causes unmagnetized blobs with higher enthalpy density to drip out of the hot plasma layer and sink into the lower enthalpy density magnetized cold plasma. These sinking blobs are subject to the RTI, which also gives rise to a \textit{secondary} Kelvin-Helmholtz instability (KHI) that can grow over a few hundreds of light-crossing times (or sound crossing times)
of the blob and shred it completely \citep[see for e.g.][for discussion of buoyant blobs in galaxy clusters, where the bubbles are shredded after few sound-crossing times of the bubble]{Reynolds+05,DS09}. The blobs are 
also accelerated downwards more slowly because of the buoyancy force that acts in the direction opposite to gravity and has magnitude 
$\mathbf F_b\approx-w_cV_b\mathbf g$, where $w_c$ is the enthalpy density of the ambient (cold) fluid and $V_b$ is the volume displaced 
by the blob. This yields a reduced downward acceleration of the blob with magnitude 
$g_{\rm eff}=(w_h-w_c)g/w_h\approx(w_{h,0}-w_{c,0})g/w_{h,0}=g/2$.

The effect of the buoyancy force and the KHI can be seen in Fig.~\ref{fig:blob} where we
show the motion of a hot unmagnetized blob in a cold magnetized fluid. The hot blob, in pressure equilibrium with the 
cold magnetized fluid, is initialized with downward velocity $v_b=0.01$ to mimic the downward dripping of hot plasma from 
the hot layer. Soon, the ram pressure exerted by the ambient fluid flattens the blob into a thin ribbon, the sides of which are 
curled upward due to the KHI. In the last snapshot of Fig.~\ref{fig:blob}, we show the velocity vectors. The KHI gives rise to 
two symmetrical vortices which cause a drag force on the stretched-out blob that largely balances the downwards pull acting on it by the effective gravity, after correcting for the buoyancy force. 

Interestingly, there is hardly any mixing between the blob and ambient fluid, which suggests that instabilities on the smallest 
scales at the interface between the blob and external fluid are suppressed. This behavior is contrary to what lower density blobs 
rising buoyantly in higher density inter-galaxy-cluster medium experience in numerical simulations. In the absence of any 
viscosity, the blobs are shredded and destroyed completely over few 
sound-crossing times. As shown by \citet{Reynolds+05,DS09}, viscosity stabilizes the secondary RTI and KHI and keeps the blob intact but 
still stretched out into a structure similar to what is shown in Fig.~\ref{fig:blob}. The simulation shown in Fig.~\ref{fig:blob} 
does not have any artificial viscosity but only that due to numerical diffusion at the grid scale, which is small but enough to suppress small scale instabilities at the interface between 
the blob and ambient fluid.

The slower motion of the hot plasma due to such instabilities has a profound effect on the rate of reconnection. As argued earlier, 
reconnection in the scenario explored here depends critically on evacuation of hot plasma from the current layer. This allows 
the cold magnetized fluids with opposite polarity to come into contact and dissipate magnetic energy. If the dripping of plasma 
out of the hot layer slows down, then so will the rate of reconnection.

\section{Discussion}\label{sec:discussion}
The ordered bulk inflow velocity $v_{\rm in}$ upstream of the current sheet is generally taken as a measure of 
the reconnection rate. In the steady-state Sweet-Parker reconnection, the inflow velocity is limited 
by the aspect ratio of the current sheet, such that $v_{\rm in} = (\delta/L) v_A$, where $\delta$ 
and $L$ are respectively the half width and half length (parallel to the upstream field direction) 
of the reconnection layer. Since in astrophysical plasmas usually $\delta/L\ll1$, the Sweet-Parker 
reconnection cannot satisfy the fast energy dissipation rates needed in any bursting phenomena. 
Alternatively, 2D and 3D MHD simulations of relativistic reconnection have shown that 
$v_{\rm in}\sim0.1v_A$ \citep[e.g.][]{WY06}, confirming the analytic result of \citet{Lyubarsky05}. 

It is important to note that these simulation are inherently different from the scenario explored here. 
In reconnection simulations similar to what is shown in \citet{WY06}, plasma is forced out of the finite simulation box along the current layer due to the magnetic tension of the reconnected field 
lines.
This allows the current layer to be evacuated at a much faster rate and yield high $v_{\rm in}$. 
In addition, such simulations also invoke an explicit finite resistivity, which greatly aids in enhancing the reconnection rate by setting up an X-type neutral point at the outset.

In this work we find turbulence rather than an ordered bulk flow in the reconnection region. Nonetheless, for comparison purposes we define $v_{\rm in}$ as the ordered bulk inflow velocity that would produce the same magnetic reconnection rate as we find in our simulations (see eq.~(\ref{eq:v_in})). With this definition we find that $10^{-3}\lesssim\beta_{\rm in}\lesssim5\times10^{-3}$, which is slower by up to two orders of magnitude. 
While $\beta_{\rm in}$ increases initially with increasing effective gravity, it quickly saturates at its 
final, but still low, value of $\beta_{\rm in}\approx 5\times10^{-3}$ (for $\sigma_0=10$) due to 
the KHI, which produces vorticity and an effective drag force that inhibits fast evacuation of hot plasma from the current layer. We solve the equations of ideal MHD without any explicit resistivity. Therefore, magnetic field diffusion is minimal (only that
due to numerical diffusion at the grid scale) which leads to most of the magnetic field away from the current 
layer to remain undisturbed. It is likely that the inclusion of finite resistivity could potentially significantly increase the reconnection 
rate.

Radiative cooling of particles in the current layer was neglected in this work, however, it can play an important role 
in determining the structure of the current layer and evolution of the KSI. If the scattering optical depth of the 
hot current layer is initially high then radiation will remain trapped inside it until photons can efficiently diffuse 
out of it. This particular scenario was explored analytically by \citet{BPL17}, where they showed that in the optically 
thick case the pressure in the current layer is dominated by the radiation field, and the width of the layer remains 
larger by many orders of magnitude as compared to the optically thin case. In the latter case, since radiation can 
stream out, loss of pressure leads to compression of the current layer and therefore a large increase in gas density, which may in turn enhance the KSI. 
Interestingly, they showed that whether radiation streams out or remains trapped in the current layer has 
no effect on the dynamics of the outflow. 

It was shown by \citet{Drenkhahn02,L10,BPL17} that a Poynting flux dominated outflow with a striped wind structure will 
accelerate due to dissipation of energy via magnetic reconnection, such that 
$d\Gamma/d\ln r\propto (\beta_{\rm in}r)^{1/3}$ for $r<r_s$, where $r_s$ is the saturation radius. The 
saturation radius $r_s\propto \Gamma_\infty^2/\beta_{\rm in}$ is the point where most of the initial magnetic energy has already been tapped so that beyond it no further acceleration 
due to magnetic reconnection is possible and the flow simply coasts at a fixed $\Gamma_\infty$. Therefore, a higher reconnection rate ($\beta_{\rm in}>0.1$) can yield a (up to an order of magnitude) lower saturation radius. As 
argued earlier, this can be facilitated by the existence of even a small resistivity in the flow.

In this work, we assumed an ideal scenario of completely anti-parallel magnetic field lines close to the current layer. 
In order to make this picture more realistic, allowance should be made for magnetic field shear between the two regions (1 and 3) with 
opposite polarity, such that there is a finite misalignment angle $\theta_{1,3}$ between the field lines. Understanding 
of the KSI in this case is important since magnetic field line tension of the misaligned field lines can stabilize the instability. 
This will be the subject of a future analytic work (Gill, Granot, \& Lyubarksy, in preparation). Exploring such a scenario numerically necessarily requires 3D simulations, which are also left for a future study.

\section*{Acknowledgements}
We would like to thank Omer Bromberg for numerous discussions on numerical MHD methods. 
RG, JG and YL acknowledge support from the Israeli Science Foundation under
Grant No. 719/14. RG is supported by an Open University of Israel Research Fund. 
All simulations in this work were carried out on an Open University of Israel 
computer cluster.



\bsp	
\label{lastpage}
\end{document}